\documentclass[structabstract]{aa}  

\usepackage{natbib}
\bibpunct{(}{)}{;}{a}{}{,} % to follow the A&A style

\usepackage{graphicx}
\usepackage{txfonts}
\usepackage{longtable}
\newcommand{\Msun}{\mbox{M$_{\sun}$}}

\def\HCO+{HCO$^+$}
\def\H13CO+{H$^{13}$CO$^+$}
\def\kms{\ifmmode {{\rm \;km\;s^{-1}}}		    	      % km s-1
       \else {\hbox{$\,${\rm km$\;$s$^{\rm -1}$}}}\fi}
\def\solar{\ifmmode_{\mathord\odot\;} \else $_{\mathord\odot}\;$\fi} % _solar
\def\mo{\ifmmode {\,{\it M}\solar\;} \else $\,M$\solar$\;$\fi}	      % M solar		
\def\cm#1{\ifmmode {\,{\rm cm^{-#1}}\;} 		              % cm-1,cm-2, cm-3, ...
	\else \hbox{$\,${\rm cm$^{\rm -#1}\;$}}\fi}
\def\am{\ifmmode{^{\scriptscriptstyle\prime}}			       % arcmin
	\else $^{\scriptscriptstyle\prime}$\fi}
\def\deg{\ifmmode{^\circ}\else$^{\circ}$\fi}			       % degree

\def\x {\ifmmode\times\else$\times$\fi}			       	       % times 

\begin{document}
   \title{A survey of the Galactic center region in HCO$^+$, H$^{13}$CO$^+$, and SiO} 
  
   \author{D. Riquelme
          \inst{1,2}
          \and
          L. Bronfman\inst{2}
	  \and
	  R. Mauersberger\inst{1,3}
          \and
	  J. May\inst{2}
	  \and
	  T. L. Wilson\inst{4}
	  }

   \offprints{D. Riquelme}

   \institute{Instituto de Radioastronom\'{i}a Milim\'etrica (IRAM),
              Av. Divina Pastora 7, Local 20, E-18012 Granada, Spain\\
              \email{riquelme@iram.es}
           \and Departamento de Astronom\'{i}a, Universidad de Chile, Casilla 36-D, Santiago, Chile
           \and Joint ALMA Observatory, Av. El Golf 40, Piso 18, Las Condes, Santiago, Chile  
           \and Naval Research Laboratory, Code 7210, Washington, DC 20375, USA
  }

   \date{Received --; accepted --}
% \abstract{}{}{}{}{} 
% 5 {} token are mandatory
 
  \abstract
   {}
   {A large-scale survey of the Galactic center region in 
  the $3$ mm rotational transitions of SiO, \HCO+ and \H13CO+ (beamsize $\sim 3\farcm 6$) was
  conducted to provide an
  estimate of cloud conditions, heating mechanisms, chemistry,
  and other properties.}
   {Using the NANTEN 4m telescope from Nagoya University, a region
  between $-5.\deg75<l<5.\deg6$ and $-0.\deg68<b<1.\deg3$ was mapped
  in the $J=1\to0$ lines of \HCO+ and \H13CO+ and in the $J=2\to1$
  line of SiO with a spacing of $3\farcm 75$ (\HCO+) and
  $1\farcm 875$ (SiO and \H13CO+).}
   {Velocity channel maps, longitude-velocity maps, and
  latitude-velocity maps are presented. We identify 51
  molecular clouds; 33 of them belong to the Galactic center and 18
  to disk gas. We derive an average of the luminosity ratio of $\rm SiO(J=2\to1)/CO(J=1\to0)$  in clouds
  belonging to the Galactic center of $4.9\times 10^{-3}$ and for disk
  clouds of $3.4\times 10^{-3}$. The luminosity ratio of $\rm HCO^+(J=1\to0)/CO(J=1\to0)$  in the Galactic center
  is $3.5\times 10^{-2}$, and for disk clouds it is $1.5\times 10^{-2}$. We can distinguish clearly between
  regions where the SiO or \HCO+ dominate.}
   {}


   \maketitle
%
%________________________________________________________________
\section{Introduction}

\indent To understand the evolution, dynamics, and constitution of our
Galaxy, it is crucial to explore its central kiloparsec. This region
is obscured by intervening dust in the optical, but not in the
millimeter to far infrared wavelength range. It contains a large
amount \citep[$\sim3\times 10^7$ {\rm \Msun},
see][]{Dahmen_et_al_1998} of molecular gas, the Central Molecular
Zone\footnote{Following the notation of \citet{Morris_Serabyn_1996},
we refer to the ``CMZ'' as the region about $-0.\deg5<l<1.\deg5$, and
to the ``Galactic center region'' as the region between
$-5\deg<l<5\deg$, which is the region observed in this
work.}\citep[CMZ,][]{Morris_Serabyn_1996}, which is traced by the
mm-emission of CO and its isotopomers
(\citealp[e.g.][]{Bitran_et_al_1997,Dahmen_et_al_1998,Bally_et_al_1987}). The
distribution and mass of the components of the interstellar medium
(ISM) in the central part of the Galaxy is discussed by
\citet{Ferriere_et_al_2007}.\\
\indent  Clouds in the Galactic center region are influenced by large
potential gradients, and the proximity to the center of our Galaxy,
which may lead to frequent cloud-cloud collisions and exposes the
clouds to enhanced magnetic fields, cosmic ray fluxes, X-rays, and
explosive events. As a consequence, in the CMZ, the lines are first,
typically wider than 10 km s$^{-1}$
\citep[e.g.][]{Morris_Serabyn_1996}. Second, the thermal emission of
SiO is extended, finding it over parsec-size regions
(\citealp[e.g.][]{Martin-Pintado_et_al_1997,Huettemeister_et_al_1998}),
which is also seen in the central regions of external galaxies
\citep{Mauersberger_Henkel_1991}. In contrast, in the Galactic disk,
SiO is observed mainly at the leading edges of outflows, which has
been interpreted as a signature of shocked gas
\citep[e.g.][]{Ziurys_et_al_1990}. In general, Galactic disk sources
are compact with sizes of $<0.1$ pc or at most 1 pc
\citep{Jimenez-Serra_et_al_2010}. Third, a substantial amount of the
gas has a kinetic temperature of $\sim 200$ K
\citep[e.g.][]{Huettemeister_et_al_1993}, while the bulk of the dust
has a much lower temperature of $T_{\rm dust}< 40$ K
(\citealp{Rodriguez-Fernandez_et_al_2002, Odenwald_Fazio_1984,
Cox_Laureijs_1989}). Our survey results provide additional information
about the heating and chemistry of Galactic center clouds that cannot
be easily obtained from an analysis of CO data alone.\\
%
%What process is heating the molecular gas? Two promising mechanisms are 
%(a) cloud-cloud collisions, and (b) ion-slip heating. The ion-slip heating mechanism refers to the
%friction between ions and neutrals, where the ionic motions are
%controlled by magnetic fields, whereas that of the neutrals by
%gravity. %Ion-slip heating will operate effectively if the fraction of
%%free electrons is very small (\citealp[see e.g.][]{Scalo_1977,
%%  Huettemeister_et_al_1993}). The presence of a large amount of the
%%formyl ion (\HCO+) may be an indicator that ion-slip heating is
%%important. 
%In the Galactic center, there is thought to be a poloidal magnetic
%field, which at the CMZ may have a strength of about 1 mG
%\citep{Morris_2006}. If this field is fixed in space,
%\citep{Morris_2006}, the Galactic center molecular clouds would move through it with
%a velocity of order 100 \kms. This can induce ionic currents, since
%molecular clouds are partially ionized by cosmic rays (CR). Ohmic
%losses of these currents can contribute the the heating of the
%gas. Little is, however, known about the details of such an ion slip
%heating in the CMZ, since magnetic fields, ionization fractions and
%the detailed geometry of the clouds and their degree of turbulence are
%poorly determined, although ionisation rates due to CR seem
%to be enhanced relative to local values by a factor of 10
%\citep{vanderTak_et_al_2006}. The origin of such CR is
%under debate, although  shock waves from supernova remnants (SNR)
%offer a viable explanation \citep{Aharonian_et_al_2006}.
%
\indent \HCO+ is a molecule known to vary
considerably in abundance relative to neutral molecules with similar
dipole moments and rotational constants, such as HCN, within a galaxy
and from galaxy to galaxy (\citealp{Nguyen_et_al_1992,
Seaquist_Frayer_2000,Krips_et_al_2008}). \citet{Seaquist_Frayer_2000}
argue that, in the environment of circumnuclear galactic or
extragalactic gas, the abundance of \HCO+ decreases with increasing CR
ionization rates. However, \citet{Krips_et_al_2008} observe that the
\HCO+ abundance tends to be higher in galaxies with nuclear starbursts
than in galaxies with active galactic nuclei (AGN), which would be
unexpected if HCO$^+$ is destroyed by the CRs produced by SNRs. Using
chemical model computations for photon-dominated regions (PDRs),
\citet{Bayet_et_al_2009} found that the molecular fractional abundance
of \HCO+ is insensitive to changes in both the CR ionization rate and
the far-UV radiation. \citet{Loenen_et_al_2008} also point out that in
PDRs the ratios of \HCO+ to HCN or HNC decrease with increasing
density and that a change in the UV flux of two orders of magnitude only
produces modest changes in the line ratio because the UV field is
attenuated at the high column densities. To provide information about
the chemistry of \HCO+ in circumnuclear regions and to relate this to
CR ionization rates and to heating mechanisms, we mapped this
molecule and its rare \H13CO+ isotopomer in its $J=1-0$ transition
throughout the Galactic center. These results can be combined with CO
data to provide insight into conditions in the Galactic center. SiO
emission, on the other hand, is a tracer of hot, shocked gas, since it
can be formed from silicon that is liberated from dust grains, either
by low-velocity shock waves or by evaporation at high temperatures.
Such shocks are expected, e.g., at the footpoints of the giant molecular
loops detected by \citet{Fukui_et_al_2006}, who explain such features
by the magnetic buoyancy caused by a Parker instability.\\
\indent Large-scale surveys have been made in $^{12}$C$^{16}$O and its
isotopomers in the $J=1\to0$, as well as in $J=2\to1$ spectral lines
(\citealp[e.g.][]{Bitran_et_al_1997, Dahmen_et_al_1998,
Oka_et_al_2001}). Up to now, there are few complete maps in species
that are less abundant than CO
(\citealp[e.g. CS:][]{Bally_et_al_1987,Bally_et_al_1988}; HNCO:
\citealt{Dahmen_et_al_1997}; NH$_3$: \citealt{Handa_et_al_2006}; OH:
\citealt{Boyce_Cohen_1994}). A compilation of existing spectral line
surveys updated from \citet{Mauersberger_Bronfman_1998} is given in
Table \ref{tablesurvey}. There have been two previous surveys of SiO
in the Galactic center region. \citet{Martin-Pintado_et_al_1997}
mapped the $J=1\to0$ spectral line, but did not cover the entire
CMZ. \citet{Huettemeister_et_al_1998} measured a number of SiO and CO
spectral line transitions toward 33 cloud maxima, to investigate the
excitation of the SiO and estimate $\rm SiO/ H_2$ ratios for the
clouds.  There is a map of the $J=1\to0$ spectral line of main
isotopic HCO$^+$ by \citet{Linke_et_al_1981}, which does not, however,
extend far beyond the Sgr A* region. \citet{Fukui_et_al_1980} also
present \HCO+ ($J=1\to0$) observations, but only toward Sgr A and a few
positions in Sgr B2.  In the following, we present maps of the
$J=2\to1$ spectral line of SiO and the $J=1\to0$ spectral lines of
HCO$^+$ and H$^{13}$CO$^+$.  These are the first complete maps of both
species in the Galactic center region.\\
\indent In Section 2 the observations and data reduction are
described. In Section 3, the survey data are presented. These consist
of the full set of the spatial maps of the integrated intensity,
longitude-velocity diagrams, and latitude-velocity diagrams for each
molecule. We present an analysis in Section 4. Four appendices are also included, beginning with Appendix A which presents complementary
figures for the paper. Also, Appendices B, C, and D present the
complete data set in \HCO+, SiO, and \H13CO+, respectively, showing
velocity channel maps of $10$ $\kms$ velocity width,
longitude-velocity, and latitude-velocity diagrams. Appendix E contains
the Gaussian fits for each cloud in the survey, identifying the
temperature peaks, velocity center, and velocity width. In a subsequent
paper, the results will be discussed in the context of other available
data.\\

\begin{table*}
\caption [] {Atomic and molecular surveys of the Galactic bulge}
\label{tablesurvey}
\begin{center} 
\begin{tabular}{cccccccc} \hline\hline
 Species & Frequ. & \multicolumn{2}{c}{Observed Area} & Sampling & FWHM &Ref.\\
         &  [GHz] & $l$ & $b$ & Interval&  & \\
         &        &[\deg]& [\deg] &  & &   \\\hline
H\ {\sc I}& 1.4    &$-11 \leq l \leq 13$& $-10 \leq b \leq 10$& $30'$ & $21'$ & $1$ \\\hline 
C\ {\sc I}& 492    &$-0.5 \leq l \leq 1.5$& $ 0 $& $3'$ & $2'$ & $2$ \\\hline
C\ {\sc II}& 1900   &$-100 \leq l \leq 60 $& $-3 \leq b \leq 3$&  & $15'$ & $3$ \\\hline
$^{12}$CO (1-0)& 115   &$-10 \leq l \leq 25$& $0^b$ & $30'^c$ &$65'$ & $4$\\\hline
$^{12}$CO (1-0)& 115   &$-12 \leq l \leq 13$& $-2 \leq b \leq 2$& $7\farcm 5^a$ &$8\farcm 8$  & $5$ \\\hline
$^{12}$CO (1-0)& 115   &$-12 \leq l \leq 12$& $-5 \leq b \leq 5$&$4'$  &$2\farcm 7$  & $6$\\\hline
$^{12}$CO (2-1)& 231   &$-3 \leq l \leq 3$& $ 0$&$7\farcm5$  &$9'$  & $7$ \\\hline
$^{12}$CO (2-1)& 231   &$-6\leq l \leq 6$ & $-2\leq b\leq 2$ &$7\farcm 5$ & $9\farcm 2$ & $8$\\\hline
$^{12}$CO (3-2)& 345   &$-1.5 < l < +1.0$& $-0.2 < b < +0.2$&   $34''$    &  $22''$ &$9$ \\\hline
$^{13}$CO (1-0)& 110   &$-5 \leq l \leq 5$& $-0.6 \leq b \leq 0.6$& $1\farcm 7$  & $6'$ & $10$ \\\hline
$^{13}$CO (1-0)& 110   &$-6 \leq l \leq 8$& $-1 \leq b \leq 1$& $2'$ & $2\farcm 6$ & $6$ \\\hline
$^{13}$CO (2-1)& 220      &$-1.5\leq l \leq 3.25$& 0 &$7\farcm 5$ &$9\farcm 2$ & $8$\\\hline
C$^{18}$O (1-0)& 110   &$-1.05 \leq l \leq 3.6$& $-0.9 \leq b \leq 0.9$& $9'$ & $9'$ & $11$ \\\hline
NH$_3$ (1,1)-(3,3)& 24 &$-2 \leq l \leq 2$& $ 0$&$1\farcm 7 $  & $5\farcm 3$ & $12$ \\\hline
SiO (1-0)& 43  &$-0.8 \leq l \leq 0.9$& $-0.25 \leq b \leq 0.2$& $2'$ & $2'$ & $13$ \\\hline
OH       & 1.7 &$-6 \leq l \leq 8$& $-1 \leq b \leq 1$& $12'$  & $10'$ & $14$ \\\hline
H$_2$CO  & 4.8 &$0.5 \leq l \leq 4$& $-0.5 \leq b \leq 0.9$& $3'$ & $3'$  & $15$ \\\hline
CS (2-1) & 98  &$-1 \leq l \leq 3.7$& $-0.4 \leq b \leq 0.4$& $1'$& $2'$  & $10$ \\\hline
HNCO(5$_{0,5}$-4$_{0,4}$)     & $110$ &$-1.05 \leq l \leq 3.6$& $-0.9 \leq b \leq 0.9$& $9'$ & $9'$  & $11$ \\\hline
HCN (1-0)& 89  &$-0.2 \leq l \leq 0.7$ & $-0.2 \leq b^d \leq 0.1$   &  $2'$ &  $2'$ & $16$ \\\hline
HCN (1-0)&  89 &$-2.15 \leq l \leq 2.15$& $ -0.3\leq b \leq 0.2$& $0\farcm 8$ & $1'$  & $17$ \\\hline
HCN (1-0)&89 &$-6 \leq l \leq  6$ & $-0.8\leq b\leq 0.87$ &$4'^e$ & $1'$ & $18$\\\hline
\end{tabular}
\end{center}
{1) \citet{Burton_Liszt_1983}, 2) \citet{Jaffe_et_al_1996}, 3) \citet{Nakagawa_et_al_1995}, 4) \citet{Bania_1986}, 5) \citet{Bitran_et_al_1997}, 6) \citet{Fukui_et_al_2006},  7)
  \citet{Oka_et_al_1996}, 8) \citet{Sawada_et_al_2001}, 9) \citet{Oka_et_al_2007}, 10) \citet{Bally_et_al_1987},  11) \citet{Dahmen_et_al_1997}, 12) \citet{Morris_et_al_1983}, 13) \citealp{Martin-Pintado_et_al_1997,Martin-Pintado_et_al_2000}, 14) \citet{Boyce_Cohen_1994}, 15) \citet{Zylka_et_al_1992}, 16) \citet{Fukui_et_al_1977}, 17) \citet{Jackson_et_al_1996}, 18) \citet{Lee_1996}, $^{\mathrm{a}}$) for $-1\deg\leq l \leq 1\deg$, $^{\mathrm{b}}$) also 4 strips in latitude at $\pm 10'$ and $\pm 20'$, $^{\mathrm{c}}$) at $b=0$, $\Delta l=6'$, $^{\mathrm{d}}$) few positions towards SgrB, $^{\mathrm{e}}$) $8'$ in regions of weak emission.
}
%\end{tabular}
\end{table*}

%__________________________________________________________________

%%%%%%%%%%%%%%%%%%%%%%%%%%%%%%%%%%%%%%%%%%%%%%%%%%%%%%%%%%%%%%%%%%%%%%%%%%%%%%%%%%
\section{Observations and data reduction}
\subsection{Observations}
%
%\begin{figure*}
%\begin{center}
%\vbox{
%%\vspace{2.0cm}
%\includegraphics[width= 0.2 \textwidth, angle=90]{figure_tex/cobertura_HCO+.eps}
%\includegraphics[width= 0.2 \textwidth, angle=90]{figure_tex/cobertura_SiO.eps}
%}
%\caption
%[]  {Spatial coverage of the observations. Top: \HCO+. Bottom: SiO and \H13CO+.}
%\label{cobertura}
%\end{center}
%\end{figure*}
\indent This survey was carried out with the NANTEN 4m telescope
operated by Nagoya University at the Las Campanas Observatory,
Chile. With its southern location and moderate angular resolution,
this instrument is well-suited to large-scale mapping of the
Galactic center region. It has a $3\farcm 5$ beamwidth at the \HCO+
frequency \citep[89.188518 GHz,][]{Lovas_et_al_1979} and a $3\farcm 6$
beamwidth at the SiO frequency \citep[86.846998
GHz,][]{Lovas_et_al_1979}, which corresponds to a spatial resolution
of about $9$ pc at a distance of $8.5$ kpc
\citep{Blitz_et_al_1993}. The front end was a 4 K cryogenically cooled
NbN superconductor-insulator-superconductor (SiS) mixer receiver that
provided a typical system temperature of $\sim 280$ K (single side
band). The spectrometer was an acousto-optical spectrometer (AOS) with
2048 channels. The frequency coverage and resolution were $250$ MHz
and 250 kHz, corresponding to a velocity coverage of 840 $\kms$ and a
velocity resolution of  $0.84$ $\kms$ at the \HCO+ frequency, and a
velocity coverage of 863 $\kms$ and a velocity resolution of $0.86$
$\kms$ at the SiO frequency. The data were calibrated using the
standard chopper-wheel method \citep{Kutner_Ulich_1981}. The measured
quantity, $T_{\rm obs}$, was converted to antenna temperature, using
$T^*_{\rm A}=T_{\rm obs}/(2\times 0.89)=T_{\rm obs}/1.78$. The factor
of 2 is needed because our raw NANTEN  data were calibrated as double
sideband, while the receiver used was single sideband.  The factor of
0.89 is a measured correction arising from the less-than-perfect image
band suppression. Throughout this work, all values are given in
$T^*_{\rm A}$. The main beam temperature scale, $T_{\rm MB}$, can be
obtained using $T_{\rm MB}=T^*_A/\eta_{\rm MB}$, where the main beam
efficiency is $\eta_{\rm MB} = 0.87$ at $86$ GHz (value provided by
the Nanten team).\\
\indent The data were observed between $1999$ and $2003$. The area
mapped is from  about  $l= -5.\deg7$ to $l= 5.\deg6$ and from $b=
-0.\deg7$ to $b= 1.\deg3$. The survey contains about $1500$ positions
in the $J =1 \to 0$ spectral line of \HCO+, at a uniform spacing
of $3\farcm 75$. Each map point was observed for at least $1$ minute
on source, for an rms noise antenna temperature of $28$ mK at a
velocity resolution of $1$ km s$^{-1}$. More than $3000$ positions
were observed in the $J= 2 \to 1$ spectral line of the vibrational
ground state of SiO, fully sampling at $1\farcm 875$ spacing  in the
most intense regions (see Fig. \ref{integrate_completo}). Each map
point was observed for $2.5$ minutes on source, for an rms noise
antenna temperature of $20$ mK at a $1$ km s$^{-1}$ velocity
resolution. The survey consisted of the CMZ, and five molecular clouds
observed in CO by \citet{Bitran_et_al_1997} with large velocity widths
that we call here  ``Peripheral Molecular Zone'' (PMZ).
% centered at the following velocities: (a)
%$(l,b,v)=(-5.\deg3, 0.\deg4, 84 $\kms$)$, Cloud 1
%\citep{Bania_1977,Bania_1986,Bania_et_al_1986}; (b)
%$(l,b,v)=(-4.\deg4,0.\deg6, 72 $\kms$)$; (c) $(l,b,v)=(-3.\deg8,
%0.\deg9,-83 $\kms$)$; (d) $(l,b,v)=(5.\deg3, -0.\deg3, 95 $\kms$)$; and a
%large velocity width cloud of 60 $\kms$  rms velocity width
%\citep{Bitran_1987} called ``M+3.2+0.3'' \citep{Bania_1977},
%$(l,b,v)=(3.\deg2, 0.\deg3, 104 $\kms$)$.
We also observed the presumably optically thin formyl ion isotopic
\H13CO+ $J =1 \to 0$ line \citep[86.754330 GHz,][]{Lovas_et_al_1979},
which is in the same spectrometer range as SiO (see Table \ref{table1}
for parameters of the survey)\footnote{The data cubes are available at http://www.das.uchile.cl/galcendata}.
\begin{table*}
\caption [] {Parameters of the Survey}
\label{table1}
\centering 
\begin{tabular}{c c c c}  
  \hline\hline                        % inserts double horizontal lines
              & \bf \HCO+ &\bf SiO & \bf \H13CO+\\\hline\hline  
Velocity resolution & 1 \kms & 1 \kms & 1 \kms\\\hline  
RMS sensitive & 28 mK & 20 mK & 20 mK\\ \hline
Spatial range: &  & &\\\hline

 CMZ & $-1.\deg5 \leq l\leq  2.\deg25$ & $-1.\deg375 \leq l\leq  2.\deg0 $& $-1.\deg375 \leq l\leq  2.\deg0$\\  &
$-0.\deg5625 \leq b\leq  0.\deg5625$ & $-0.\deg5 \leq b\leq  0.\deg5$ & $-0.\deg5 \leq b\leq  0.\deg5$ \\\hline
 M+3.2+0.3 (Clump 2) & $2.\deg5625 \leq l\leq  3.\deg5 $ & $2.\deg5625 \leq l\leq  3.\deg4375$ & $2.\deg5625 \leq l\leq  3.\deg4375$\\
                 & $-0.\deg3125 \leq b\leq  0.\deg8125$ & $-0.\deg25 \leq b\leq  0.\deg875 $   & $-0.\deg25 \leq b\leq  0.\deg875$\\\hline
 M$-$5.3+0.4  & $-5.\deg75 \leq l\leq  -4.\deg6875 $ & $-5.\deg8125 \leq l\leq -4.\deg6875$ & $-5.\deg8125\leq l\leq -4.\deg6875$\\
                 & $-0.\deg125 \leq b\leq  0.\deg5625 $ & $-0.\deg125 \leq b\leq  0.\deg5625$ & $-0.\deg125 \leq b\leq  0.\deg5625$\\ \hline
 M$-$4.4+0.6      & $-4.\deg75 \leq l\leq  -4.\deg25$  & $-4.\deg75 \leq  l\leq  -4.\deg3125$  & $-4.\deg75 \leq l\leq  -4.\deg3125$\\
                 & $0.\deg25 \leq b\leq  0.\deg8125$  & $0.\deg1875  \leq b\leq  0.\deg84375$ & $0.\deg1875 \leq b\leq  0.\deg84375$\\\hline
 M$-$3.8+0.9      & $-4.\deg0 \leq l\leq  -3.\deg625 $   & $-4.\deg0625  \leq l\leq  -3.\deg625$ & $-4.\deg0625 \leq l\leq  -3.\deg625$\\
                 & $0.\deg5625 \leq b\leq  1.\deg1875 $  & $0.\deg5625  \leq b\leq  1.\deg21875$ & $0.\deg5625 \leq b\leq  1.\deg21875$\\\hline
 M+5.3$-$0.3      & $5.\deg125 \leq l\leq  5.\deg625$   &  $5.\deg0625 \leq l\leq  5.\deg6875$& $5.\deg0625 \leq l\leq  5.\deg6875$\\
                 & $-0.\deg6875 \leq b\leq  0.\deg125 $ & $-0.\deg8125 \leq b\leq  0.\deg125$ & $-0.\deg8125 \leq b\leq  0.\deg125$\\\hline

 Sampling interval:& $3\farcm 75$  & $3\farcm 75$ & $3\farcm 75$\\
                 &              &  $1\farcm 875^{\mathrm{a}}$& $ 1\farcm 875^{\mathrm{a}}$ \\\hline
 Velocity range: & & & \\\hline
 CMZ      & $-350\kms \leq {\rm v}\leq  350\kms $ & $-280\kms \leq {\rm v}\leq  300\kms $ & $-300\kms \leq {\rm v}\leq  250\kms  $\\ \hline
 M+3.2+0.3  & $-300\kms \leq {\rm v}\leq  300\kms $ & $-150\kms \leq {\rm v}\leq  350\kms $ & $-180\kms \leq {\rm v}\leq  200\kms  $ \\\hline
 M$-$5.3+0.4  & $-350\kms \leq {\rm v}\leq  350\kms $ & $-150\kms \leq {\rm v}\leq  210\kms $ & $-190\kms \leq {\rm v}\leq  260\kms  $ \\ \hline
 M$-$4.4+0.6  & $-350\kms \leq {\rm v}\leq  350\kms $ & $-150\kms \leq {\rm v}\leq  210\kms $ & $-190\kms \leq {\rm v}\leq  260\kms  $ \\ \hline
 M$-$3.8+0.9  & $-350\kms \leq {\rm v}\leq  350\kms $ & $-320\kms \leq {\rm v}\leq  190\kms $ & $-190\kms \leq {\rm v}\leq  130\kms  $ \\ \hline
 M+5.3$-$0.3  & $-300\kms \leq {\rm v}\leq  300\kms $ & $-100\kms \leq {\rm v}\leq  270\kms $ & $-190\kms \leq {\rm v}\leq 280\kms  $ \\ \hline
\end{tabular}
\begin{list}{}{}
\item[$^{\mathrm{a}}$] in the most intense zones, see
  Fig. \ref{integrate_completo}.
\item M+3.2+0.3, M$-$5.3+0.4, M$-$4.4+0.6 , M$-$3.8+0.9, M+5.3$-$0.3,
  are the clouds observed by \citet{Bitran_et_al_1997} and the positions
  of the cloud are defined by them.  
\end{list}
\end{table*}
%__________________________________________________________________
\subsection{Data reduction}
\indent The data were reduced using the NDRS (Nanten Data Reduction
Software) package. Each data point was reduced individually, fixing the
velocity emission interval, and the order of the polynomial to fit the
spectrum baseline. Most baselines were polynomials of first order, but
a few spectra required second and third order baselines to produce
flat spectra where no spectral line emission was expected. To determine the
emission interval, we used the velocity intervals where CO($1 \to 0$)
emission appears in the Galactic center \citep{Bitran_et_al_1997}.\\
\indent For the reduction of spectra belonging to the CMZ and
M+3.2+0.3 cloud, we used the longitude-velocity diagrams of CO
\citep{Bitran_et_al_1997} as a guide to determine the range of
possible molecular emission required for baseline subtraction. The
reduction and evaluation of \HCO+ data was straightforward since this
spectral line is not blended with other strong molecular emission. The
SiO reduction was more difficult since within the SiO band, spectra
are offset by +320 $\kms$ from the \H13CO+ $(1\to0)$  spectral
line. Given the large linewidths present in the CMZ, these spectra are
sometimes nearly blended. These two spectral lines were reduced
independently. In most of the spectra, the SiO and \H13CO+ emission
are  clearly separated, but toward $l<-0\deg.5 $ in CMZ, the SiO
spectrum shows high-velocity emission ($\sim 170$ km s$^{-1}$, see
Fig. 4 middle), and a priori is not clear  whether this emission
corresponds to a high velocity clouds in SiO, or to a low velocity
cloud in \H13CO+ ( $\sim-150$ \kms). In most cases, a comparison with
(unblended) main isotopic \HCO+ could settle this ambiguity. For
very few spectra ($\sim$ 10 spectra), we had to study adjacent
spectra to distinguish between emission from SiO and \H13CO+. This
kind of problem was only found in the CMZ in longitudes lower than
$-0.\deg5$. In most cases ($\sim 90\%$), the emission appears to come
from high-velocity SiO cloud, rather than \H13CO+.\\
\indent In the PMZ, the data reduction was more difficult owing to the
low signal-to-noise ratios in the spectra ($\sim 5$ and even less in
M$-$3.8+0.9 cloud in \H13CO+) and the high linewidths. % and, in the case of SiO and \H13CO+, line blending.  
In these cases, we used the CO data from \citet{Bitran_1987} to  define 
the velocity interval range where the emission is possible. We first 
obtained a summed spectrum over the total cloud and from that defined 
the velocity range of emission. To establish the baseline, we subsequently 
reduced each spectrum individually. To reduce the \HCO+ spectra we interpolated the
CO data, which have a sampling of $7\farcm 5$, to the same grid as the
\HCO+ ($3\farcm 75$), and then compared spectrum by spectrum. For SiO, we
interpolated the \HCO+ data to the SiO sampling ($1\farcm 875$), and
finally, for \H13CO+, we  used \HCO+ and SiO data to define the
velocity range where emission might be present. The polynomial order
of the subtracted baselines was typically higher than for the
CMZ. Most spectra required a polynomial order below 3, but in a
few cases, we had to use fourth grade. The basic result of the survey
are data cubes, i.e.only,  spectra obtained point by point in longitude and
latitude, forming a three-dimensional array of ${T^*_{\rm A}}$. We
obtained three data sets with coordinates galactic longitude-galactic
latitude-radial velocity for each observed molecule.
%
%%%%%%%%%%%%%%%%%%%%%%%%%%%%%%%%%%%%%%%%%%%%%%%%%%%%%%%%%%%%%%%%%%%%%%%%%%%%%%%%%%%%%%%
\section{Results}
\begin{figure*}
\begin{center}
\hbox{
\includegraphics[width= 0.35 \textwidth, angle=90]{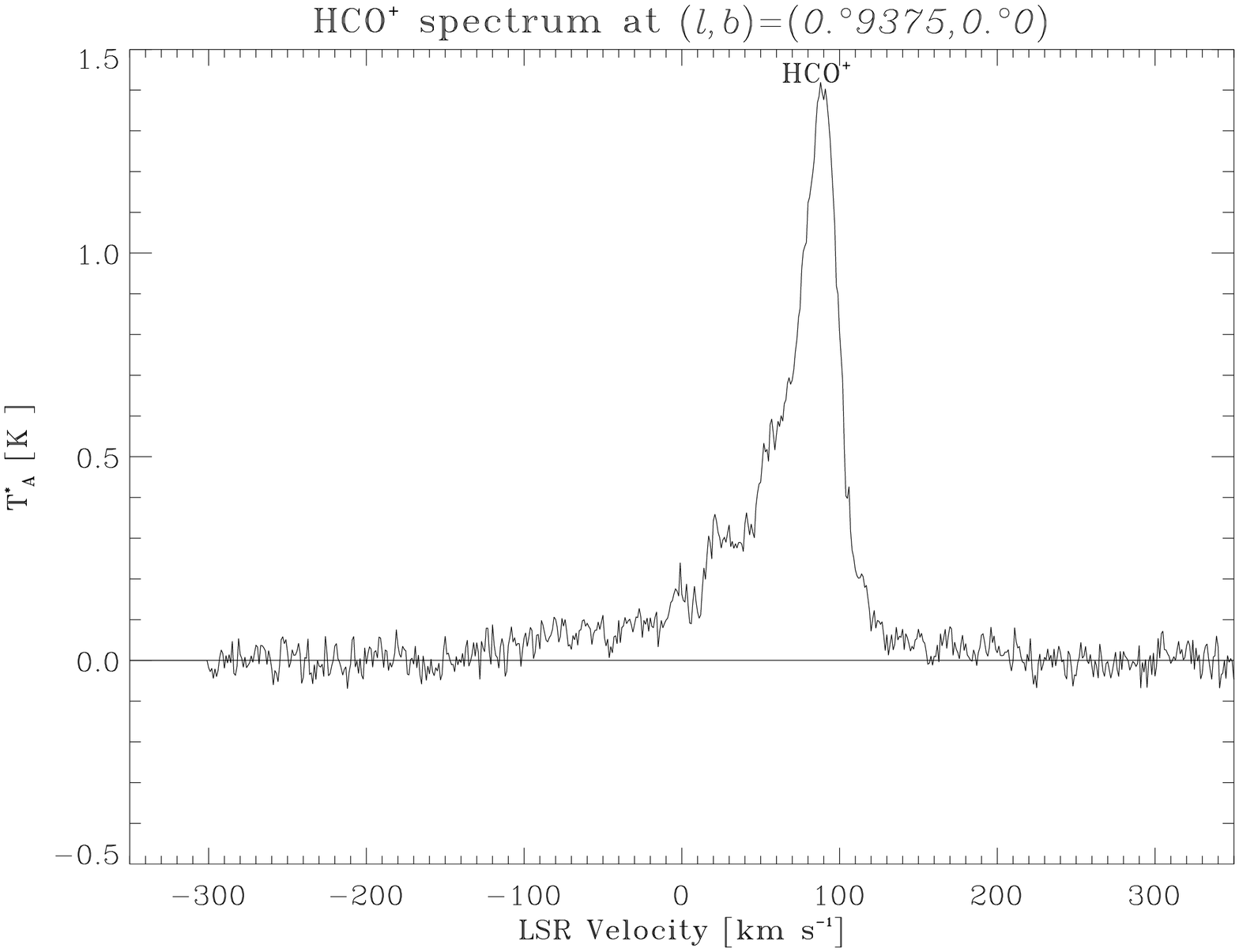}
\includegraphics[width= 0.35\textwidth, angle=90]{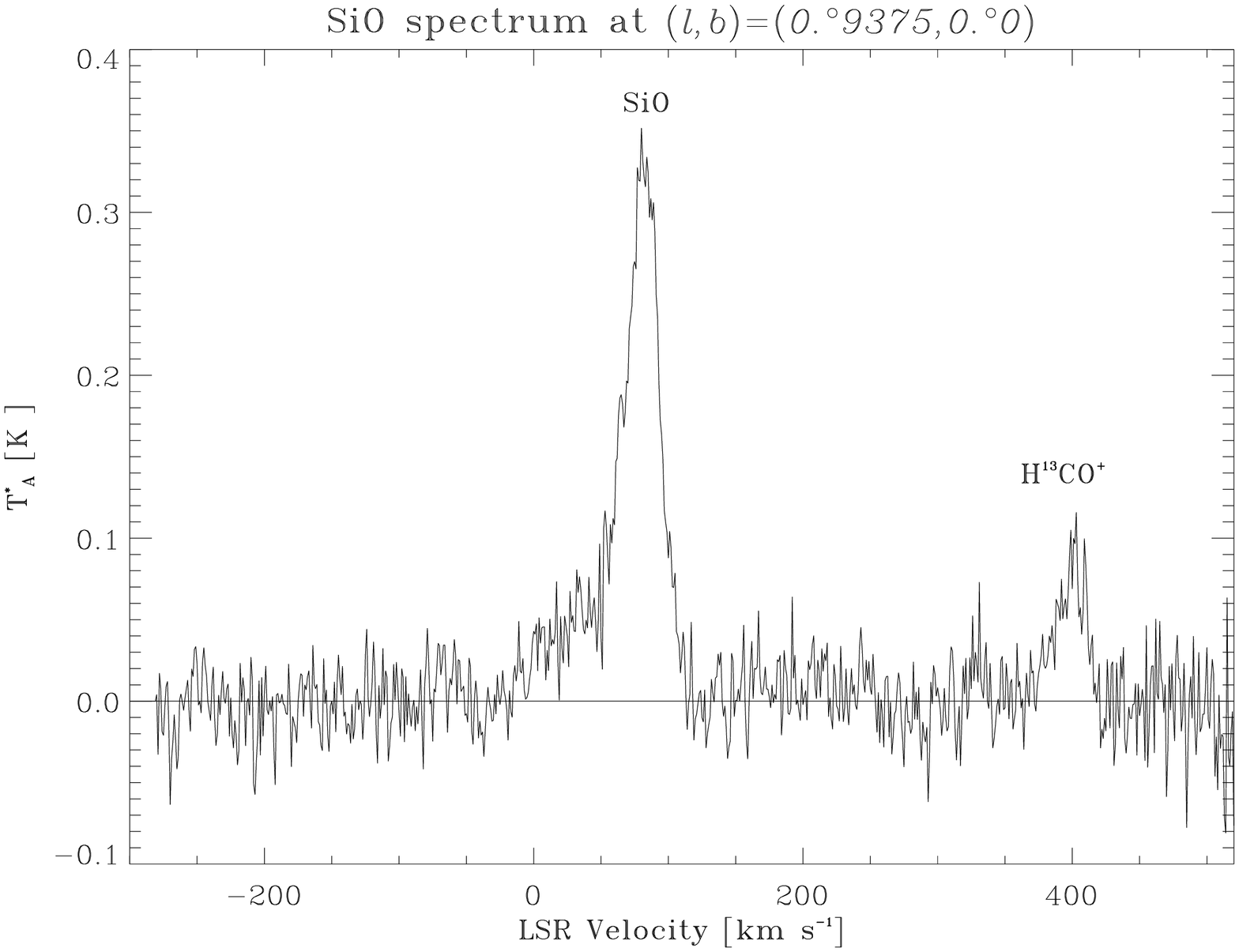}
}
\caption
[]  {Typical spectra of \HCO+ (left) and SiO and \H13CO+ (right) at $l$ and $b$ $(0.\deg9375,0.\deg0)$.}
\label{espectro_HCO}
\end{center}
\end{figure*}

\indent In this section we present the results of the \HCO+ $(1\to0)$,
SiO $(2\to1)$, and \H13CO+$(1\to0)$ Galactic center
survey. Figure \ref{espectro_HCO} shows typical spectra of \HCO+,
\H13CO+, and SiO. The \HCO+ spectrum shows emission
over a very broad velocity range between -150 \kms and 100 \kms.
%
%__________________________________________________________________________________________
\subsection{The integrated intensity maps}
\begin{figure*}
\begin{center}
\vbox{
\includegraphics[width=0.18\textwidth,angle=90]{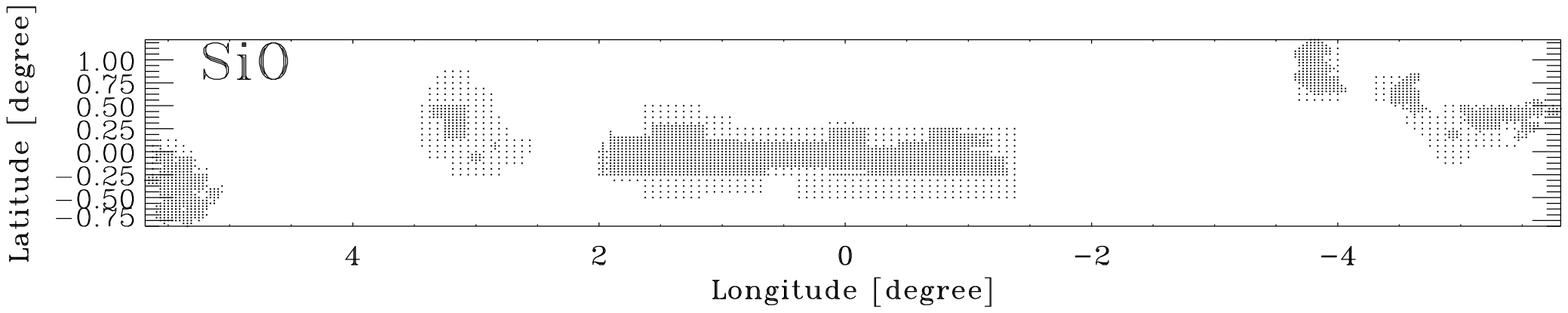}
%\vspace{-0.5cm}
\includegraphics[width=0.18\textwidth,angle=90]{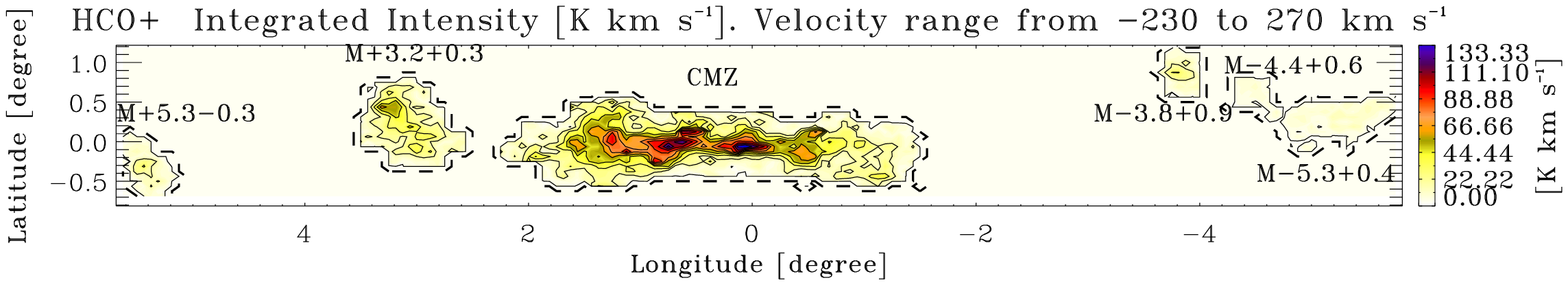}
%\vspace{-0.5cm}
\includegraphics[width=0.18\textwidth,angle=90]{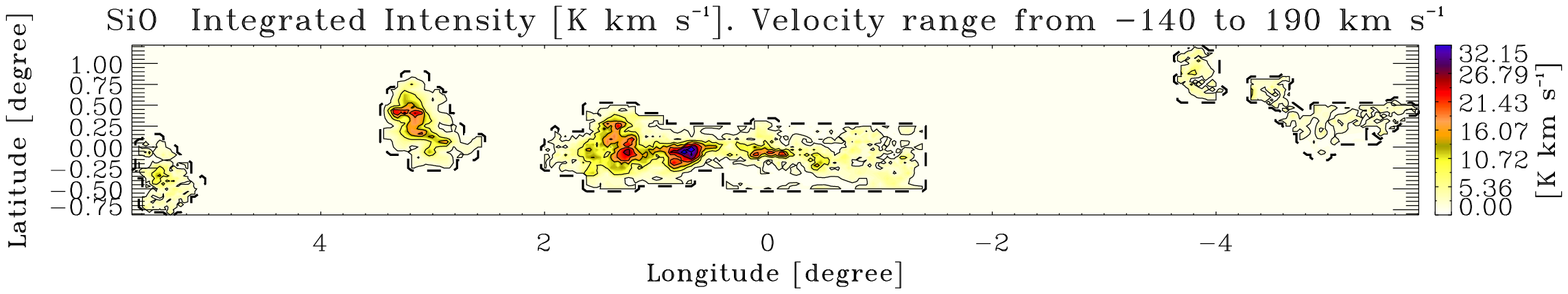}
%\vspace{-0.5cm}
\includegraphics[width=0.18\textwidth,angle=90]{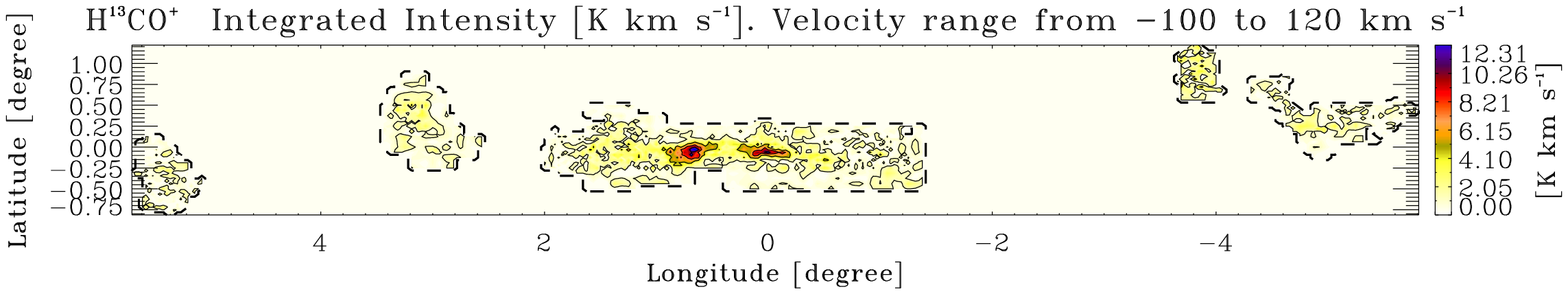}
}
\caption[]{From top to bottom: Spatial coverage of the observations in SiO and \H13CO+ (\HCO+ has uniform sampling of $3\farcm 5$).\\ Emission integrated over velocity from $-230$ to $270$
  $\kms$ for the region measured in \HCO+. The solid
contour levels start at $1.9$ K $\kms$(3$\sigma$) and increase in
steps of 12.53 K $\kms$ (20$\sigma$). \\Emission integrated over velocity from $-140$ to $190$
  $\kms$ for the region measured in SiO. The solid contours start at $1.09$ K $\kms$ ($3\sigma$)
  and increase in steps of $5.46$ K $\kms$(15$\sigma$). \\
  Emission integrated over velocity from $-100$ to $120$
  $\kms$ for the region measured in \H13CO+. The solid contours start at $0.89$ K $\kms$ ($3\sigma$)
  and increase in steps of $2.97$ K $\kms$ ($10\sigma$).\\ In all plots, the dashed line shows the coverage of the survey in each molecule. For a better display of the
 observations, we choose the velocity integration range in each spectral line to cover only the
 emission visible in the respective longitude-velocity diagram}
\label{integrate_completo}
\end{center}
\end{figure*}
\indent In Fig. \ref{integrate_completo}, we show the integrated
intensity maps, $\int T^*_{\rm A} {\rm dv}$, of the entire observed
region in the \HCO+, SiO, and \H13CO+ spectral lines. For a better
display of the observations, we chose the velocity integration range
in each spectral line to cover only the emission visible in the
respective longitude-velocity diagram, which is indicated in the
figure captions. The lowest contour level is at $3\sigma$. The value
of $\sigma$ was calculated as
\begin{equation}
\label{1sigma_lb}
\sigma_{lb}= \sqrt{N_{\rm v}}\times \Delta {\rm v} \times T^*_{\rm A}(rms),
\end{equation}
where $N_{\rm v}$ is the number of velocity channels covered by the
emission (for example, in \HCO+, the emission is within $-230$ to
$270$ \kms, therefore $N_{\rm v}=501$), $\Delta \rm v$ is the velocity
resolution (1 \kms), and T$^*_{\rm A}(\rm rms)$ is $28$ mK for \HCO+
and 20 mK for SiO and \H13CO+. In Fig. \ref{integrate_completo}, we
can distinguish both the CMZ and the PMZ.  In the SiO and \H13CO+
maps, the spacing of the observations was variable ($1\farcm 875$ in
the most intense regions and $3\farcm 75$ for the remaining of the
maps, see top of Fig. \ref{integrate_completo}). We therefore
interpolated the map to the positions with no observations.\\
\indent For an easy comparison with previous work, most of which include only the CMZ and the cloud at $l\sim3\deg.2$, in
Appendix A, we plot the integrated intensity emission only in this
region in all spectral lines observed
(Fig. A.1). In this figure we can see the
well-known asymmetry of molecular distribution with respect to the
Galactic center, with the emission concentrated on the positive
longitude side (e.g. \citealp{Sawada_et_al_2001,Oka_et_al_1996}), and
the broad features of the CMZ such as Sgr A ($l\sim 0\deg$), Sgr B
($l\sim 0.\deg6$), Sgr C ($l\sim -0.\deg5$), Sgr D ($l\sim 1.\deg1$),
Sgr E ($l\sim -1.\deg1$, $v~-200$ $\kms$, see
e.g. \citealp{Liszt_2006}), and the $1.\deg3$ complex. In the \HCO+
spectral line, both in the velocity-integrated map and in the channel
maps (Appendix B), the most intense source corresponds to
$l=0.\deg0625$ and $b=-0.\deg0625$ (in Sgr A region). In SiO emission,
the intensity peak of the whole map is at $l=0.\deg75$ and
$b=-0.\deg0625$, i.e. the Sgr B2 region. In \H13CO+, the highest
intensity is toward Sgr B and Sgr A; the other CMZ features are less
intense. In the Appendix B.1, C.1, and D.1, we present integrated
intensity maps  in these spectral lines in $10$ $\kms$ wide velocity
intervals.
%
%__________________________________________________________________________________________
\subsection{Longitude-velocity plots}
\begin{figure*}
\begin{center}
\vbox{
\includegraphics[width=0.4\textwidth,angle=90]{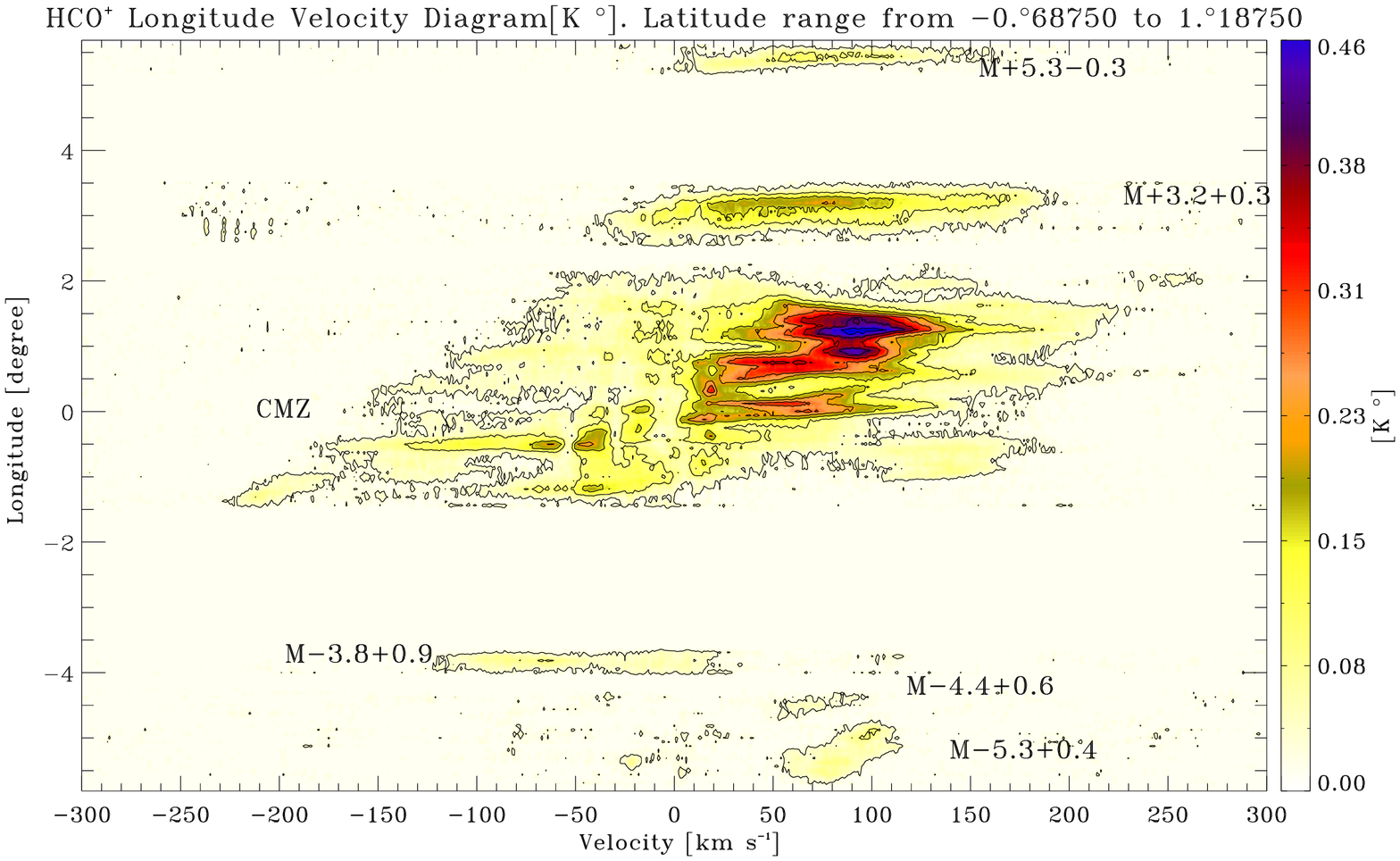}
%\hspace{-1.0cm}
\includegraphics[width=0.4\textwidth,angle=90]{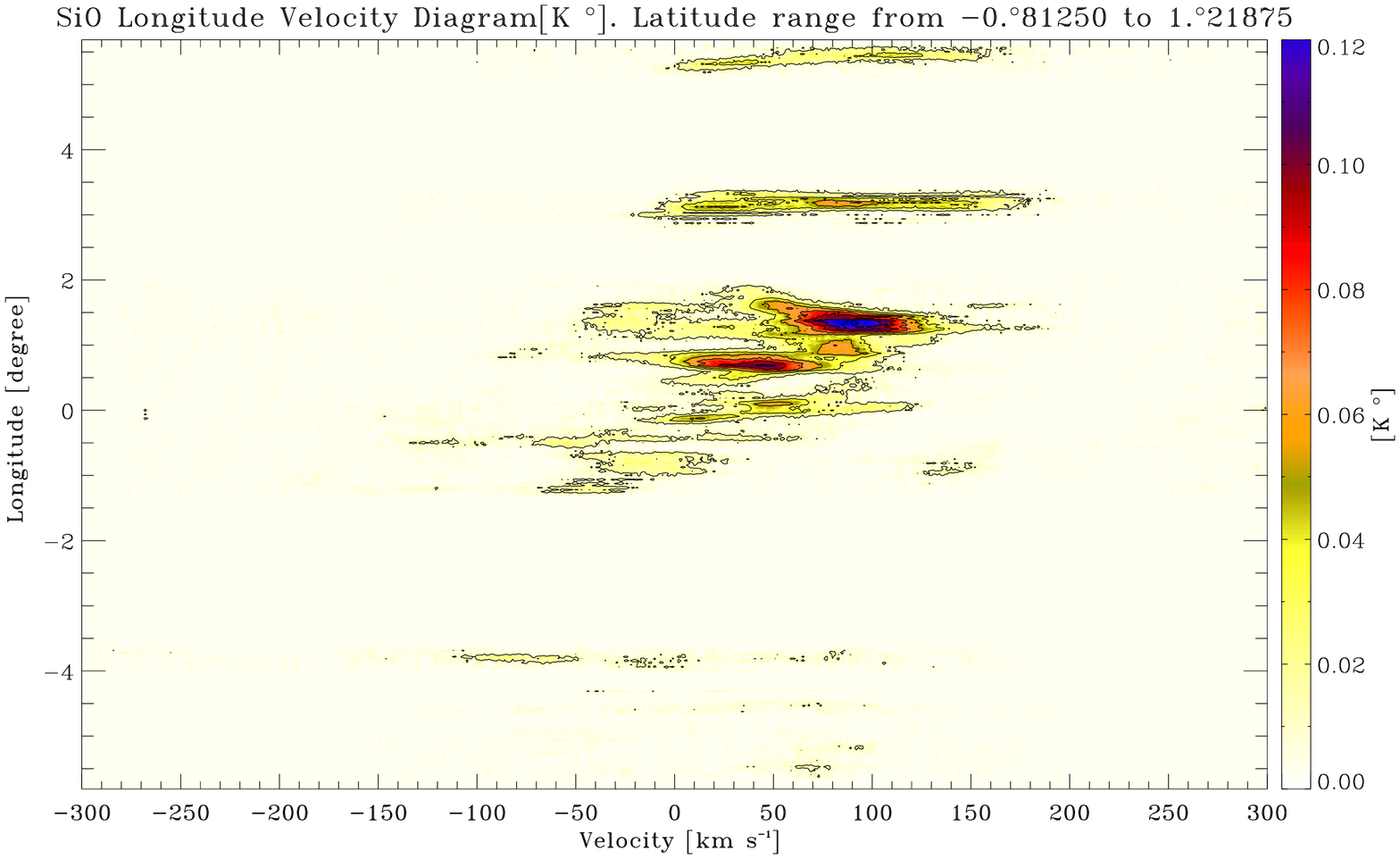}
%\hspace{-1.0cm}
\includegraphics[width=0.4\textwidth,angle=90]{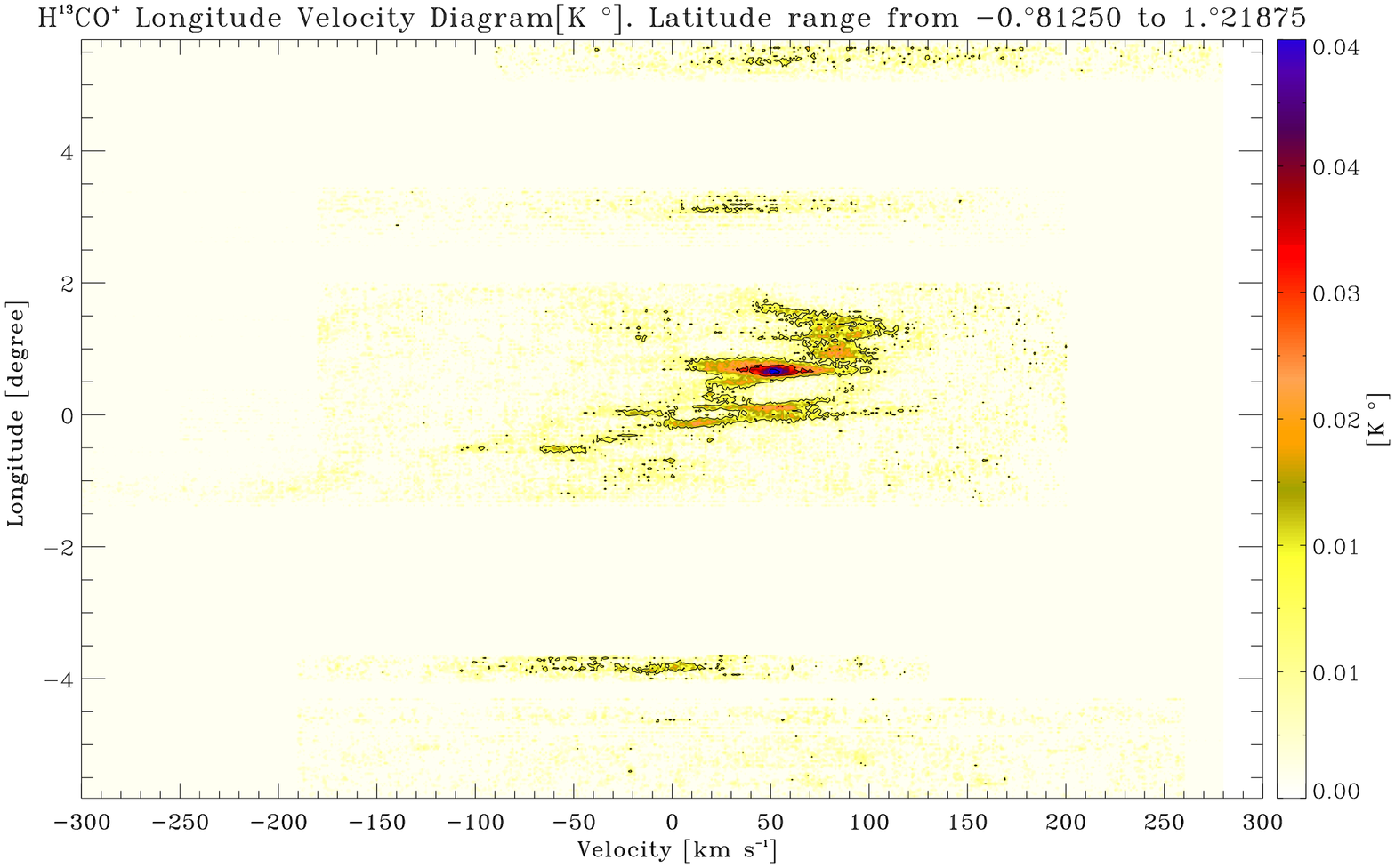}
}
\caption[]  {Top: Longitude-velocity diagram of \HCO+ emission from
  the CMZ and PMZ covering the whole survey in the latitude range
  between $b=-0.\deg6875$ to $b=1.\deg1875$. The contour levels start
  at $0.021$ K\deg ($3\sigma$), and increase in steps of $0.058$ K\deg
  (8$\sigma$).\\ Middle: Longitude-velocity diagram of SiO emission
  from the CMZ and PMZ covering the whole survey in the latitude range
  between $b=-0.\deg8125$ to $b=1.\deg21875$. The contour levels start
  at $0.01$ K\deg ($3\sigma$), and increase in steps of $0.018$ K\deg
  ($5\sigma$).\\ Bottom: Longitude-velocity diagram of \H13CO+
  emission from the CMZ and PMZ covering the  whole survey in the
  latitude range between $b=-0.\deg8125$ to $b=1.\deg21875$. The
  contour levels start at $0.009$ K\deg ($3\sigma$), and increase in
  steps of $0.016$ K\deg ($5\sigma$).}
\label{longitud_velocidad_todo}
\end{center}
\end{figure*}
\begin{figure*}
\begin{center}
\vbox{
\vbox{
\includegraphics[width=0.25\textwidth,angle=90]{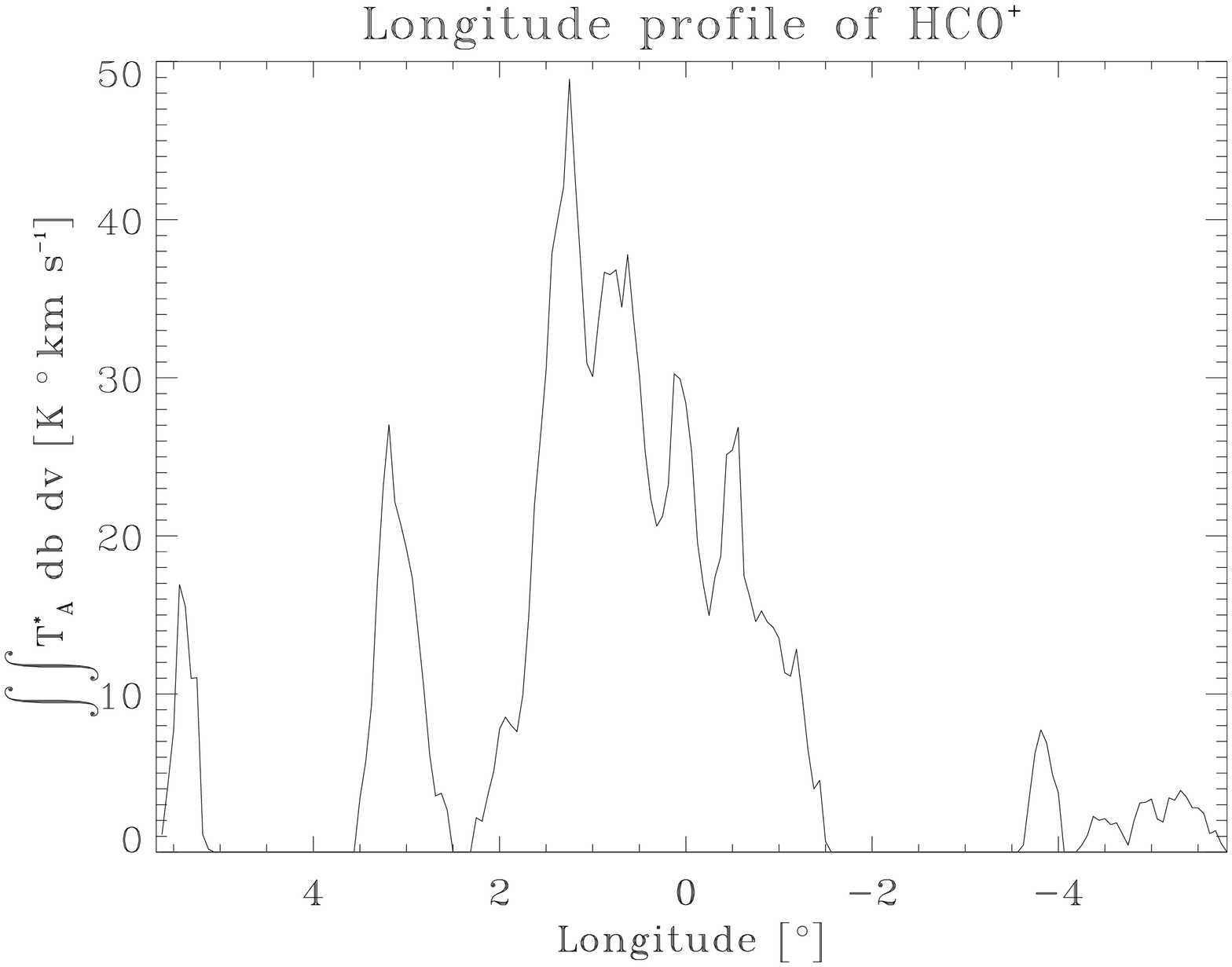}
\includegraphics[width=0.25\textwidth,angle=90]{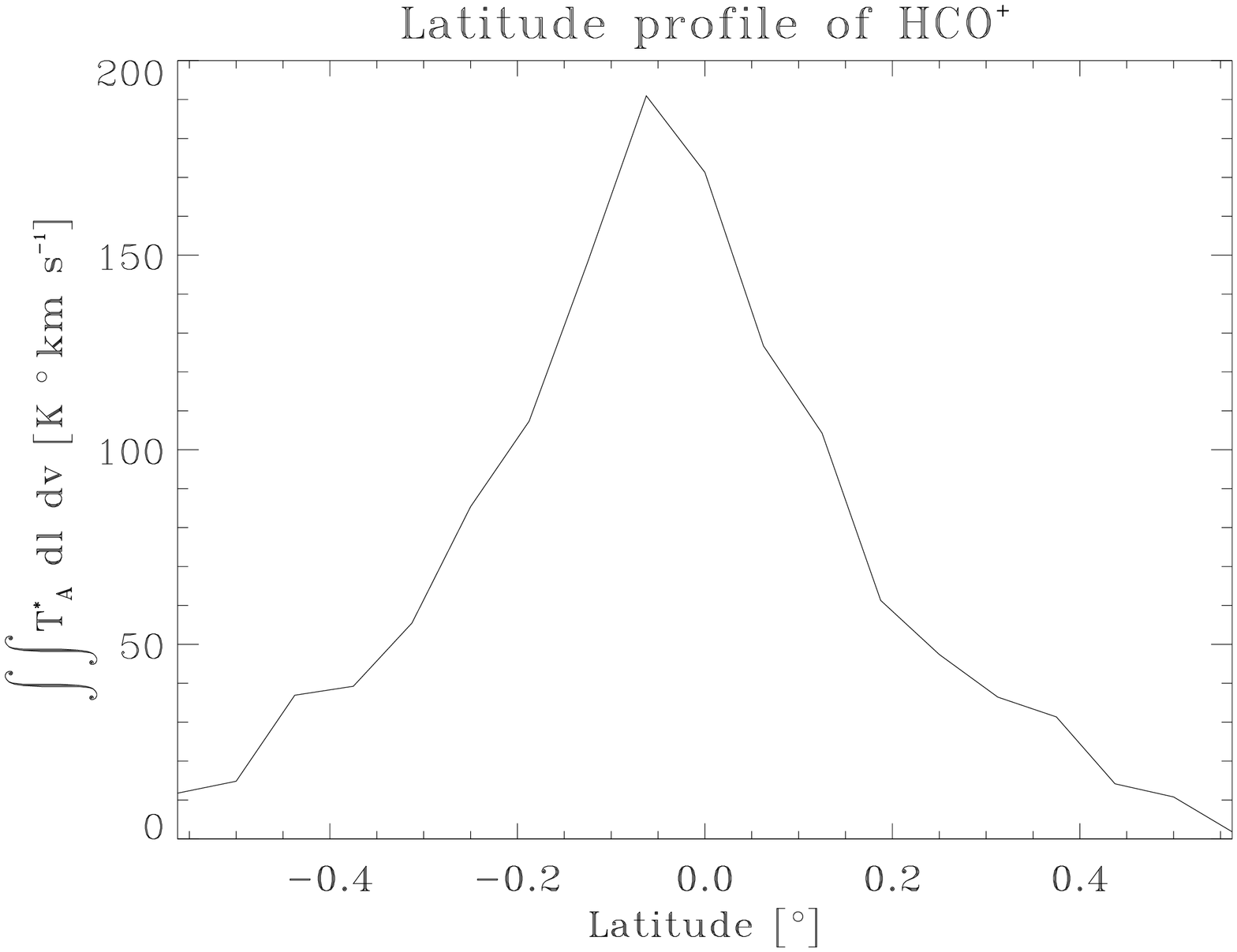}
}
\vbox{
\includegraphics[width=0.25\textwidth,angle=90]{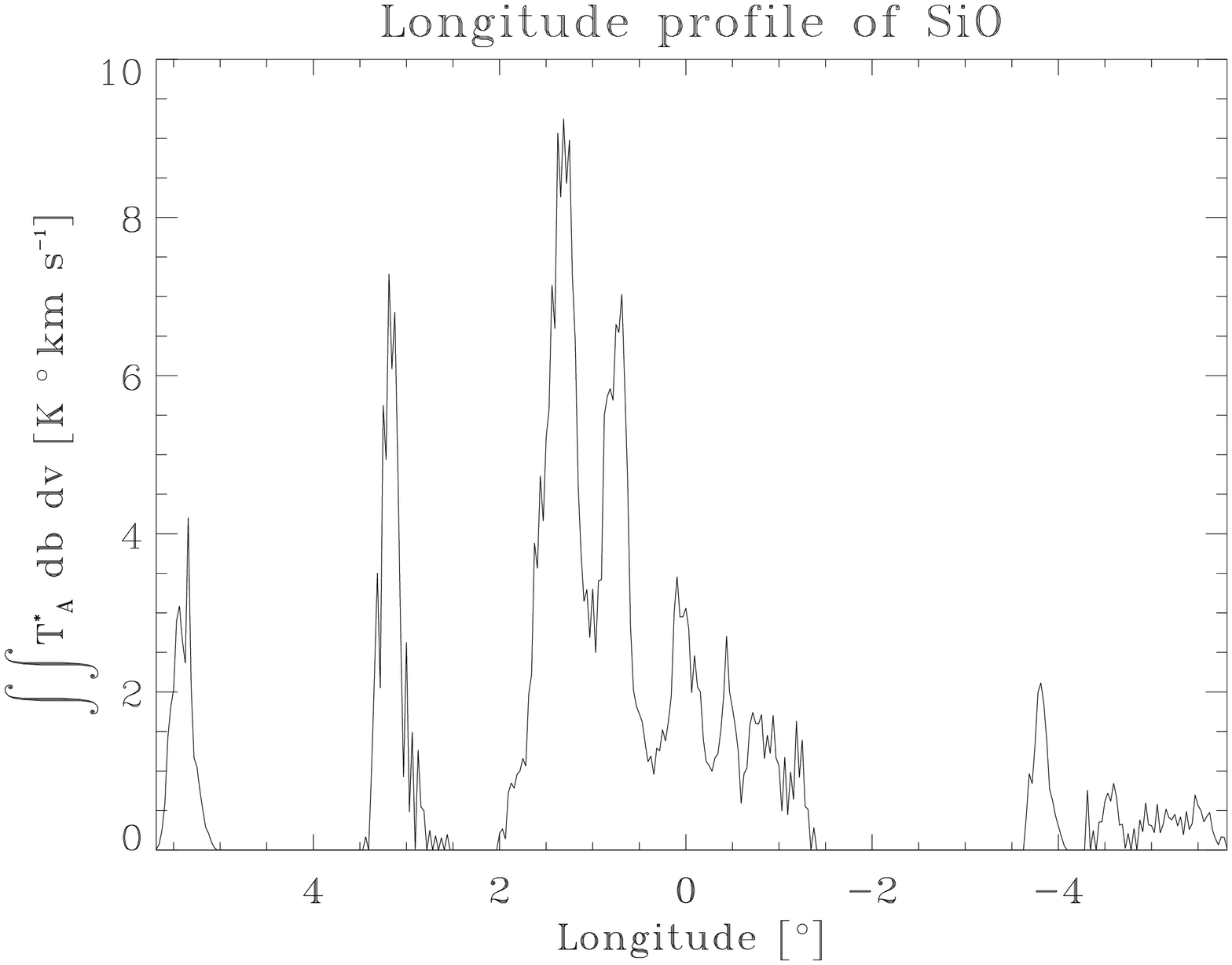}
\includegraphics[width=0.25\textwidth,angle=90]{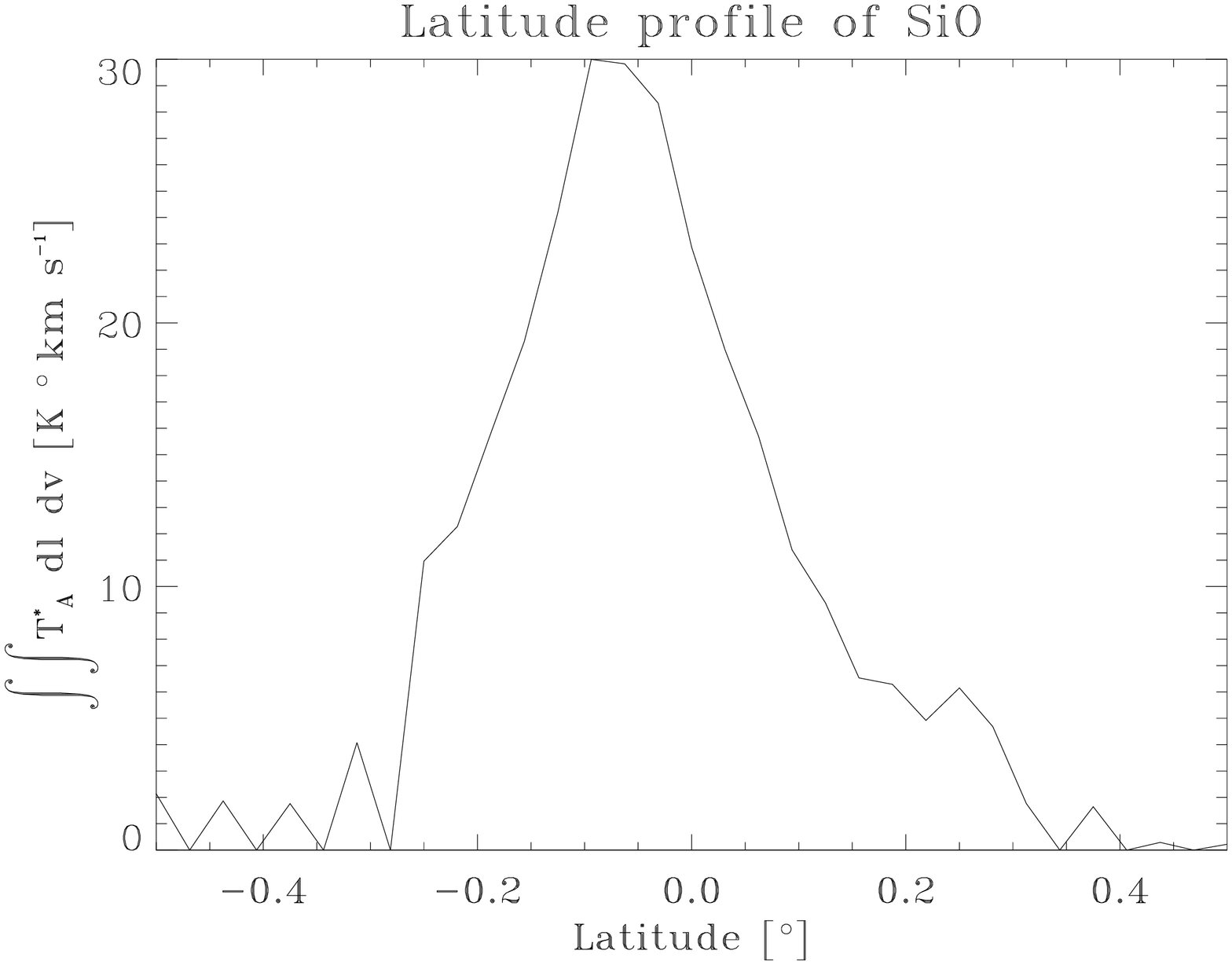}
}
\vbox{
\includegraphics[width=0.25\textwidth,angle=90]{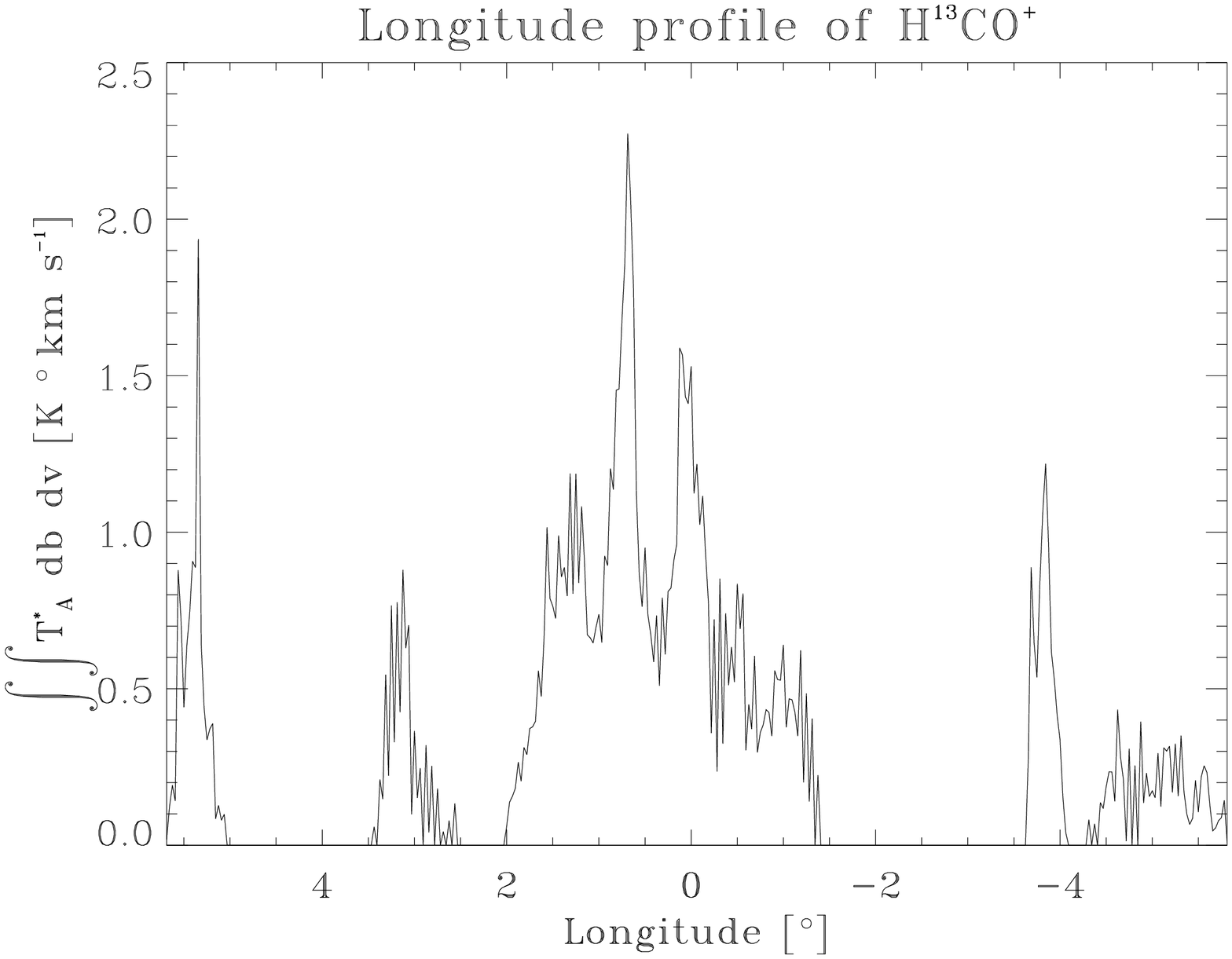}
\includegraphics[width=0.25\textwidth,angle=90]{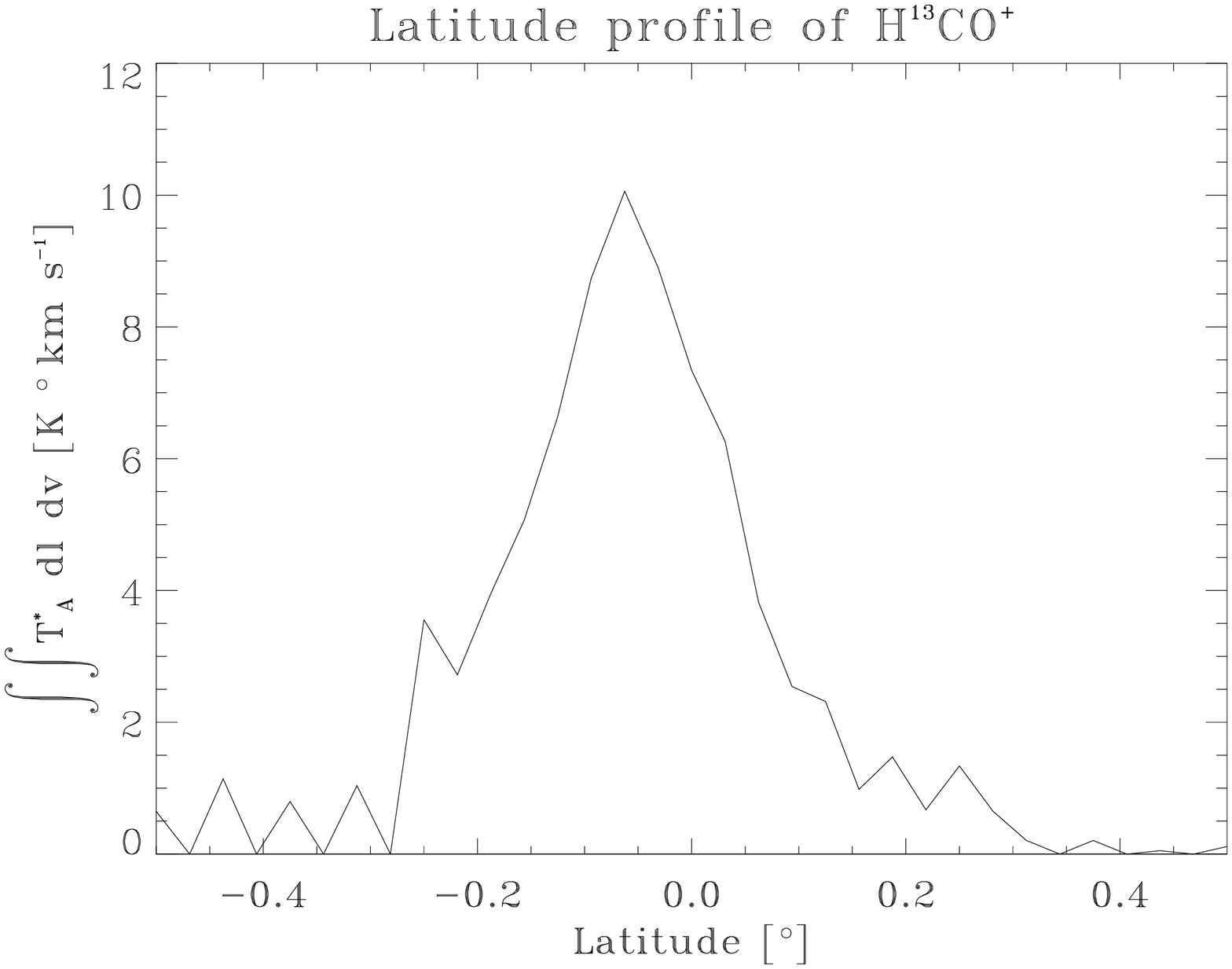}
}
}
\caption[]{Left: longitude profile for the entire latitude range observed in \HCO+ (top), SiO (middle) and \H13CO+ (bottom) emission. Right:  Latitude profile for entire longitude range observed of the
CMZ ($-1.\deg5<l<2.\deg25$) in \HCO+ (top), SiO (middle) and \H13CO+ (bottom) emission.}
\label{perfiles}
\end{center}
\end{figure*} 
\indent In Fig. \ref{longitud_velocidad_todo}, we plot the 
intensity integrated in latitude for the survey ($\int T^*_{\rm A} {\rm d}b$),
covering all the observed range as a function of $l$. In \HCO+ and SiO maps,
we can clearly see the CMZ and the PMZ, which appear
as broad features. In \H13CO+, the CMZ is weak as seen, and the only cloud
clearly seen is the M-3.8+0.9. The lowest level of the contours is
$3\sigma$, which was calculated using
\begin{equation}\label{1sigma}
\sigma_{l\rm v}= \sqrt{N_b}\times \Delta b\times T^*_{\rm A}(\rm rms),
\end {equation}
where $N_b$ is the number of latitude points with latitude
emission (e.g., 17 pixels for \HCO+ in the CMZ), and $\Delta b$ the
spacing in latitude ($0.\deg0625$ for \HCO+ data and $0.\deg03125$ for
SiO and \H13CO+ data).\\
\indent In Appendix A, we plot the integrated intensity in latitude for
the CMZ and M+3.2+0.3 cloud (Fig. A.2). We
see the well known asymmetry in longitude and velocity as a 
parallelogram shape, with the emission placed primarily at
positive velocities for $l>0$\deg and at negative velocities for
$l<0$\deg. We see the large molecular complex features, such as Sgr C
($l \sim -0.\deg7$ to $-0.\deg1$; $v<0$ \kms), Sgr A \citep[$l \sim 0$\deg,
$v\sim 50\kms$;][]{Fukui_et_al_1977}, Sgr B ($l\sim 0.\deg6$,$\rm
v\sim 50\kms$), and Sgr D ($l \sim 0.\deg9$, $v\sim 80$ \kms) and the
$1.\deg3$ complex, with a strong peak  at $l\sim 1.\deg25$ and $v\sim
90$\kms. The molecular gas complex associated to Sgr E is barely seen
toward  $l \sim -1.1$\deg, $v\sim -200\kms$ \citep[e.g.,][]{Liszt_2006}.
As already noted in previous surveys of CO and H{\sc I}
(e.g. \citealp{Bitran_et_al_1997,Burton_Liszt_1983}), the molecular
gas at the Galactic center shows non-circular movements with
velocities forbidden for galactic rotation, negative for $l>0$\deg,
and positive for $l<0$\deg.\\
\indent In the \HCO+ map, the foreground spiral arms appear as narrow
absorption features at $l=0\deg$ with ${v_{\rm LSR}}\sim -50$ km
s$^{-1}$ (3 kpc arm), ${v_{\rm LSR}}\sim -30$ km s$^{-1}$ (Norma
arm), and ${v_{\rm LSR}}\sim 0$ km s$^{-1}$ (Crux am). These
absorption features were previously observed in \HCO+ and HCN
(\citealp{Fukui_et_al_1977,Fukui_et_al_1980,Linke_et_al_1981}). In SiO we do not detect any 
absorption. The SiO emission appears to be more fragmented
than \HCO+. Sgr E is weaker than other features. At $l\sim
0$\deg, we detect the well-known clouds associated with Sgr A. Compared with the others features,
Sgr B is very intense. The 1.\deg3 complex is the most intense feature
in this map. In SiO, there is less emission with forbidden velocities
than in \HCO+. One example of a region where it is not immediately clear
whether the emission arises from high-velocity (forbidden) SiO or from
\H13CO+ can be seen in the mid panel of Fig. \ref{longitud_velocidad_todo}
 (or with more detail in Fig. A.2) at $l\sim -0.\deg8$ and
$v \sim 150$ \kms. A comparison with the unblended \HCO+ emission (top
panel) clearly suggests that this feature arises from a forbidden
velocity component of SiO. In \H13CO+, we can see the features of the CMZ,
but these are much weaker than in the other molecular lines. Sgr C
shows a very weak emission, and we can barely detect the cloud at
$l=-0.\deg5$ and $v=-50$ \kms. Sgr A is more intense, and one can see three
clouds at $(l,v)\sim (0$\deg$,-15$\kms), $(l,v)\sim (-0.\deg125,
15$\kms), and at $(l,v)\sim (0.125$\deg$, 50$\kms). The last two
could correspond to the molecular complex  M-0.13-0.08 y
M-0.02-0.07, with velocities of $+20$ $\kms$ and $+50$ $\kms$,
respectively \citep{Martin-Pintado_et_al_1997}. Sgr B2 is the most
intense feature, with an intensity peak at
$l=0.\deg65625$ and $v= 50$ \kms. Sgr D is less intense than in the
other molecular lines, and the 1\deg.3 complex is very weak. In this spectral line,
M+3.2+0.3 is barely visible. In Appendix B.2, C.2, and D.2, we present a
set of longitude-velocity diagrams, one for each observed latitude in
\HCO+, SiO, and \H13CO+.\\
\indent We show the longitudinal distribution of the molecular emission in Fig. 
\ref{perfiles}, $I(l)=\int \int T^*_A {\rm d}b{\rm d}v$, integrated
over the whole observed latitude. We can see that, in the longitude corresponding to the CMZ, the
emission appears asymmetrically distributed toward $l>0\deg$, and the
$5$ clouds in the PMZ clearly appear as intensity peaks at $l\sim 3$\deg,
$5.\deg5$, $-3.\deg8$, $-4.\deg4$, and $-5.\deg3$. In the CMZ, most of
the emission is found toward $l>0$\deg, obtaining an average longitude
weighted by intensity of $0.\deg5$ for \HCO+, $0.\deg7$ for SiO, and $0.\deg4$
for \H13CO+.
%__________________________________________________________________________________________
\subsection{Latitude-velocity plots}
\begin{figure*}
\begin{center}
%\hspace{-1.5cm}
\vbox{
\includegraphics[width=0.3\textwidth,angle=90]{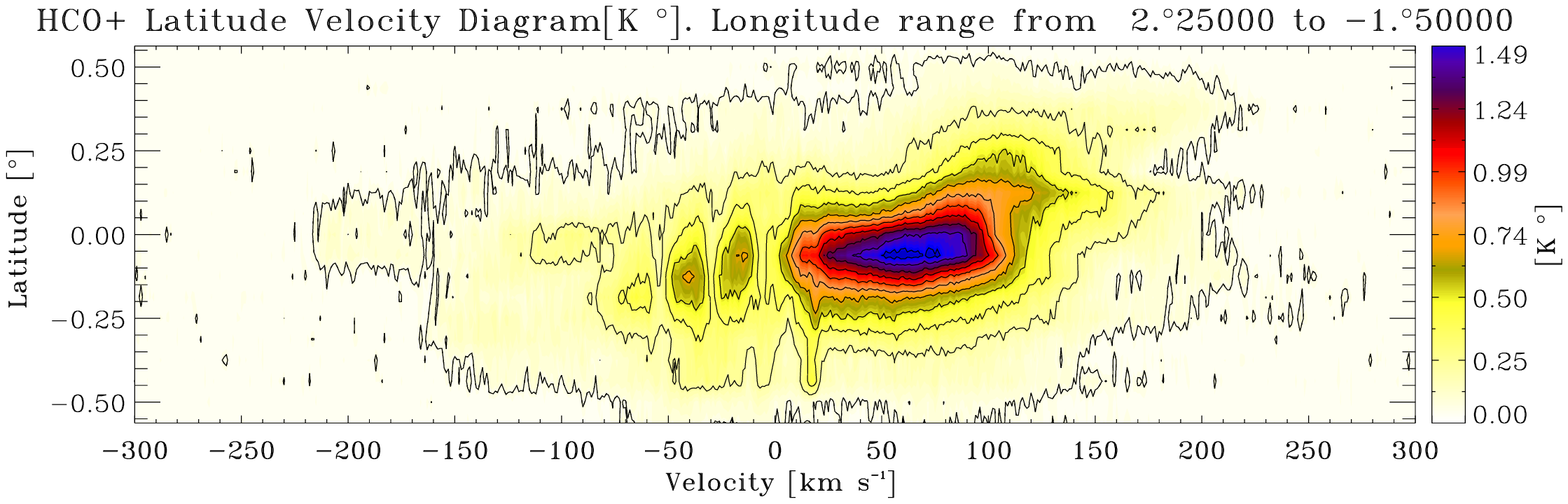}
\includegraphics[width=0.3\textwidth,angle=90]{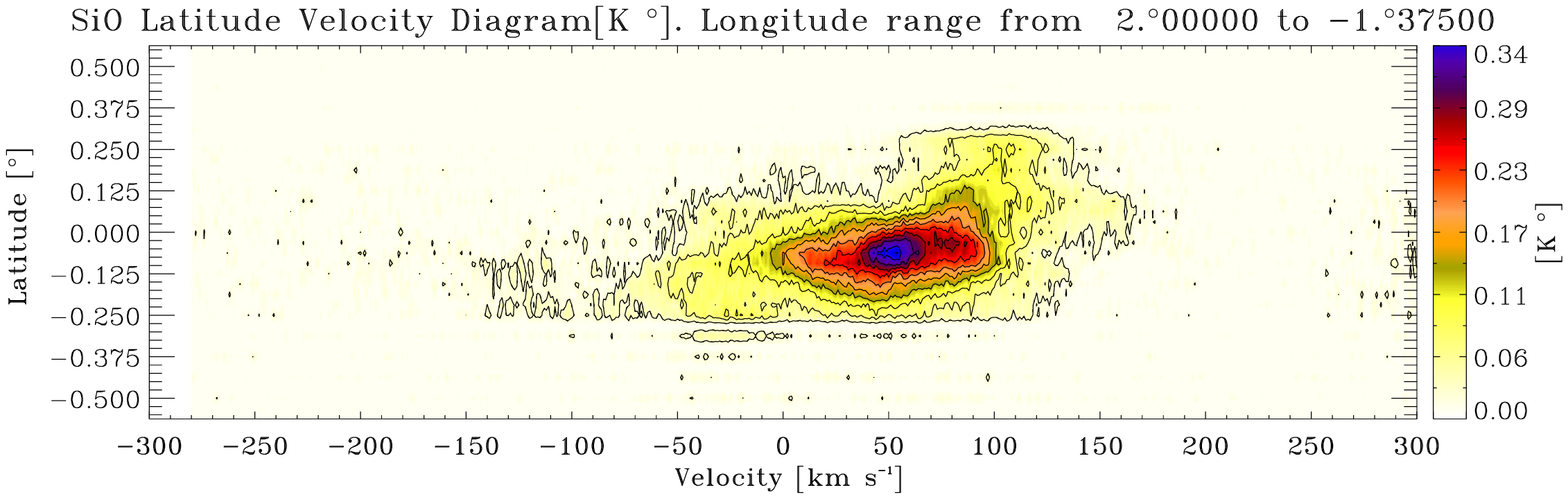}
}
\caption[]
{Latitude-velocity diagram of the CMZ. Top: \HCO+. The contours start
at $0.041$ K\deg ($3\sigma$ value) and increase in steps of $0.136$
K\deg ($10\sigma$). Bottom: SiO. The contours start
at $0.02$ K\deg ($3\sigma$ value) and increase in steps of $0.03$
K\deg ($5\sigma$).}
\label{latitud_velocidad_CMZ}
\end{center}
\end{figure*}
\begin{figure*}
\begin{center}
\hspace{-1.5cm}
\vbox{
\includegraphics[width=0.3\textwidth,angle=90]{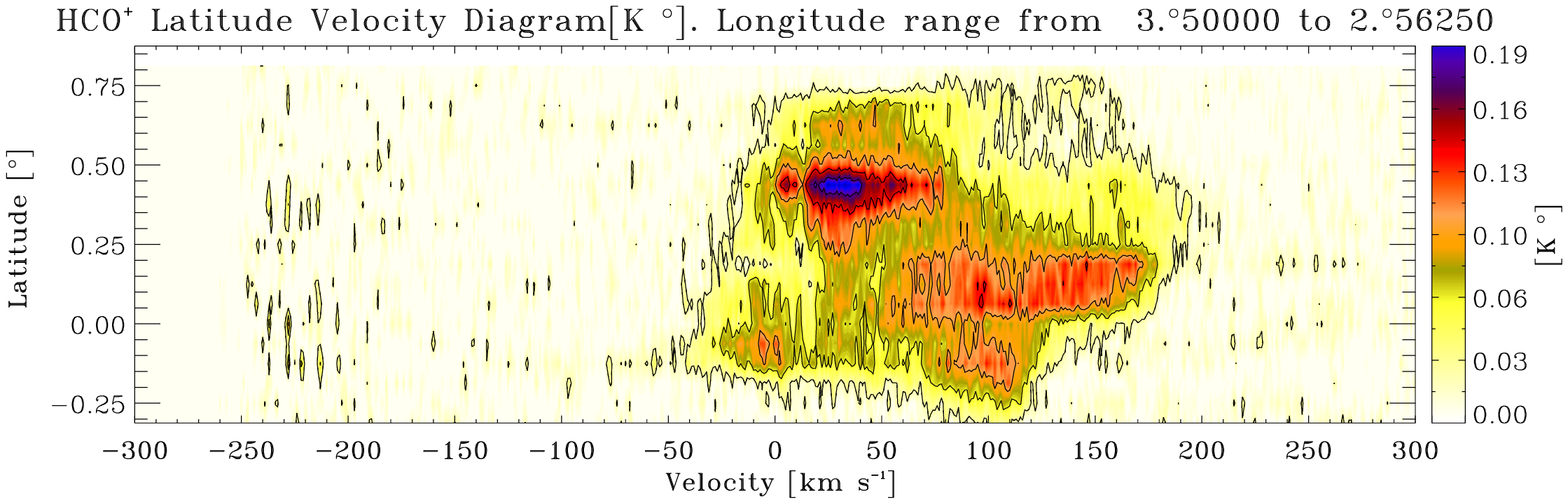}
\includegraphics[width=0.3\textwidth,angle=90]{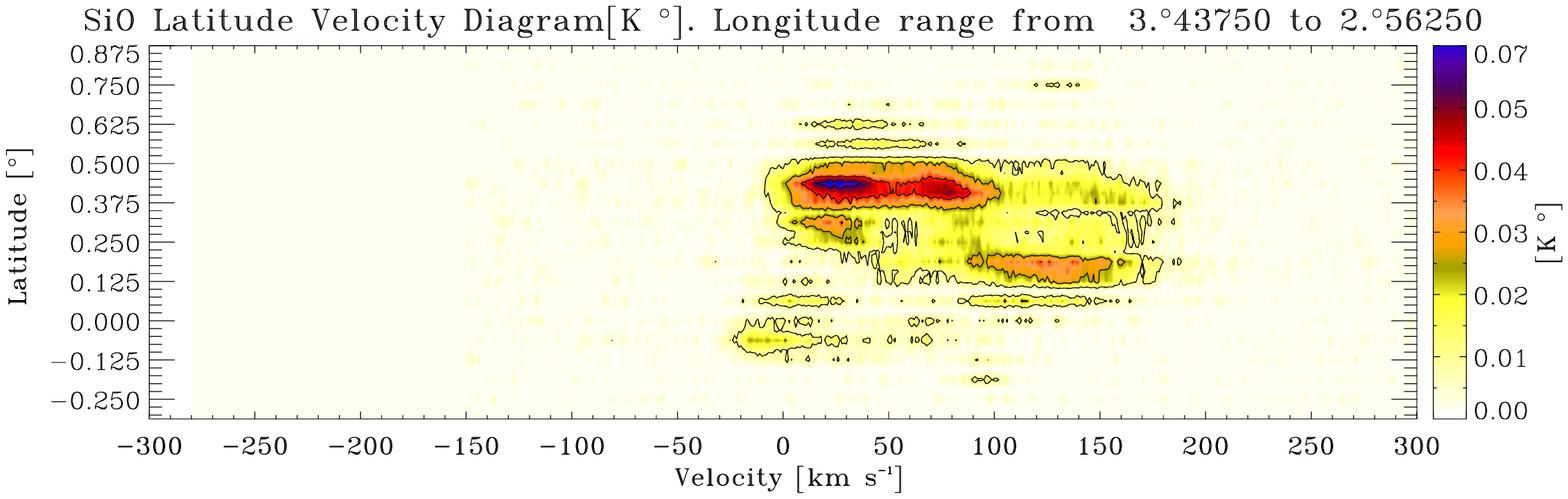}
}
\caption[]
{Latitude-velocity diagram of M+3.2+0.3. Top: \HCO+. The contours start at $0.021$ K\deg ($3\sigma$ value) and increase in
steps of $0.035$ K\deg ($5\sigma$). Bottom: SiO. The contours start
at $0.009$ K\deg ($3\sigma$ value) and increase in 
steps of $0.015$ K\deg ($5\sigma$).}
\label{latitud_velocidad_clump2}
\end{center}
\end{figure*}
\indent We present the intensity integrated in longitude ($\int
T^*_{\rm A} {\rm d}l$), covering all observed range for the CMZ and
M+3.2+0.3 cloud.  Figure \ref{latitud_velocidad_CMZ} shows the CMZ,
integrated in all observed longitude corresponding to this region. In
Fig. \ref{latitud_velocidad_clump2}, we show the integrated intensity
in longitude from \HCO+, SiO, and \H13CO+ for the M+3.2+0.3 cloud. In
the \HCO+ map, the absorption produced by the spiral arms in $\sim
-50$ \kms, $\sim -30$$\kms$, and $\sim 0$ \kms is apparent.\\
\indent We present the latitude profile of the CMZ, $I(b)=\int \int
T^*_{\rm A} {\rm d}l{\rm d}v$ (Fig. \ref{perfiles}), integrated over
the entire observed longitude. The average latitude weighted by
intensity in \HCO+ emission is $-0.\deg04$ (which agrees with the
value of $-0.\deg05$ obtained by \citealt{Bitran_1987} for CO emission),
in SiO, $-0.\deg04$, and in \H13CO+ is $-0.\deg06$.\\
\indent In the Appendices B.3, C.3, and D.3 we show the latitude-velocity
diagrams, one for each observed longitude.
%
%%%%%%%%%%%%%%%%%%%%%%%%%%%%%%%%%%%%%%%%%%%%%%%%%%%%%%%%%%%%%%%%%%%%
%
\section{Discussion}
As mentioned in the previous section, all molecules observed by us are
widely distributed throughout the Galactic center region. To
distinguish between the dominant heating mechanism for the molecular gas in well-determined space and velocity regions, we compare the maps of SiO and
\HCO+ and the maps of SiO and \H13CO+.
\subsection{The spatial and velocity distributions of the spectral line emission ratios}
\begin{figure*}
\begin{center}
\vbox{
\includegraphics[width=0.3\textwidth,angle=90]{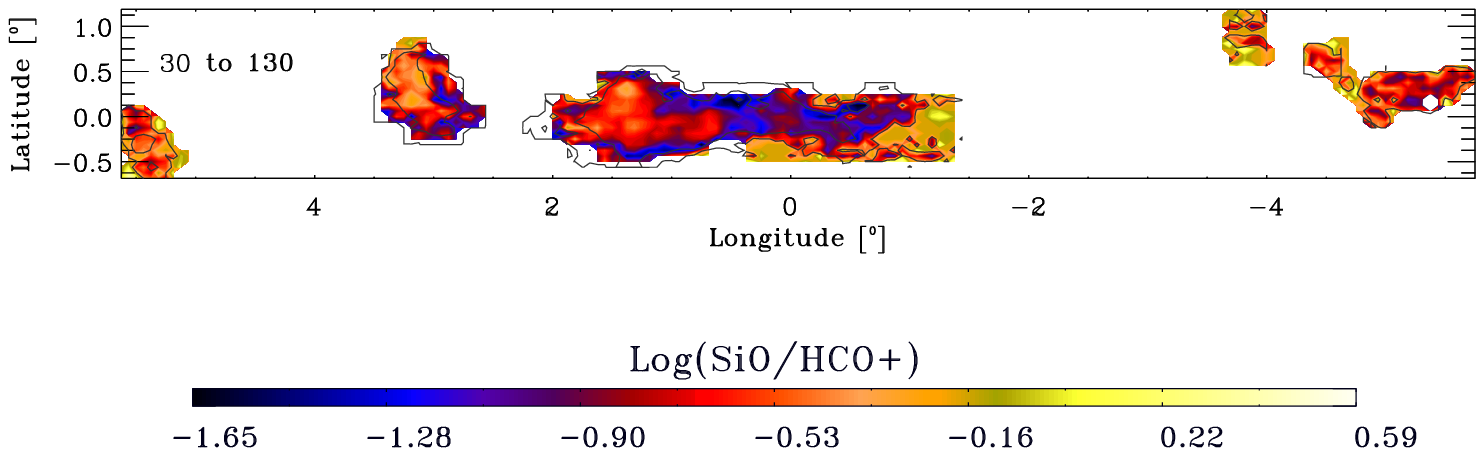}
\includegraphics[width=0.3\textwidth,angle=90]{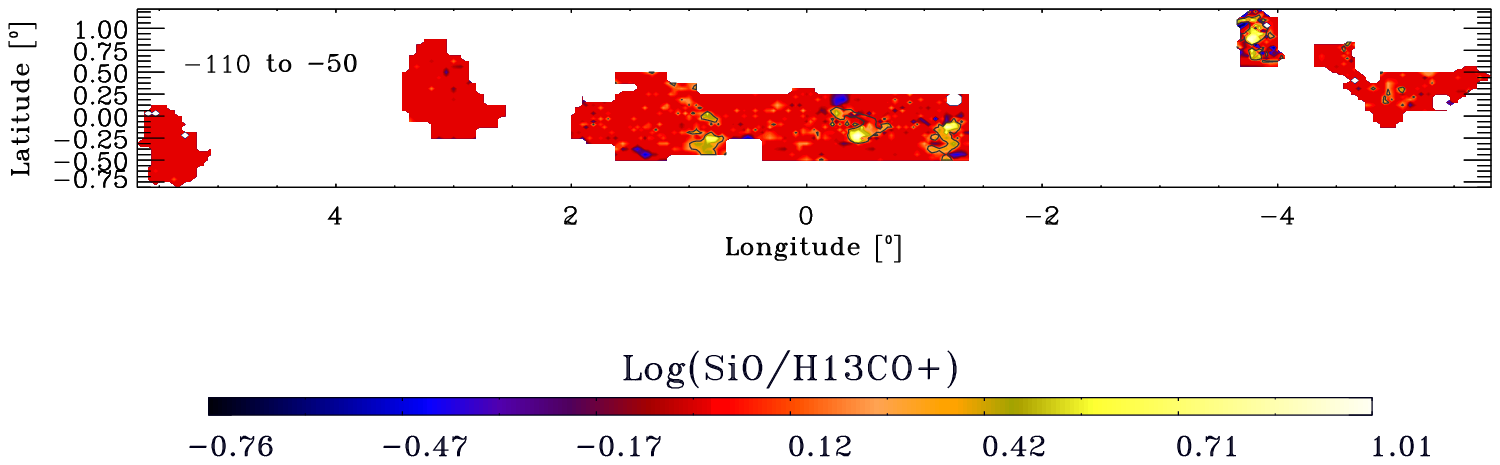}
}
\caption{Logarithm of the integrated intensity ratio. Top: $\rm
  log(\int T(SiO) dv/ \int T(HCO^+) dv)$ in the velocity range from 30
  to 130 \kms. The contours correspond to the \HCO+ emission at $3
  \sigma$ and $30 \sigma$. We can identify clearly regions where
  either the SiO (e.g. in the 1.3 Complex and in the M+3.2+0.3 cloud)
  or \HCO+ (e.g. towards Sgr A region) dominate. Bottom: $\rm log(\int
  T(SiO) dv/\int T(H^{13}CO+) dv)$ in the velocity range from $-110$
  to $-50$ \kms. The contours correspond to the SiO emission at $3
  \sigma$ and $30 \sigma$. In this velocity range, we can see the
  enhacement of the SiO toward the M$-$3.8+0.9 cloud.}
\label{comparacion_espacial_moleculas}
\end{center}
\end{figure*}
\begin{figure*}
\begin{center}
\includegraphics[width=1.0\textwidth,angle=0]{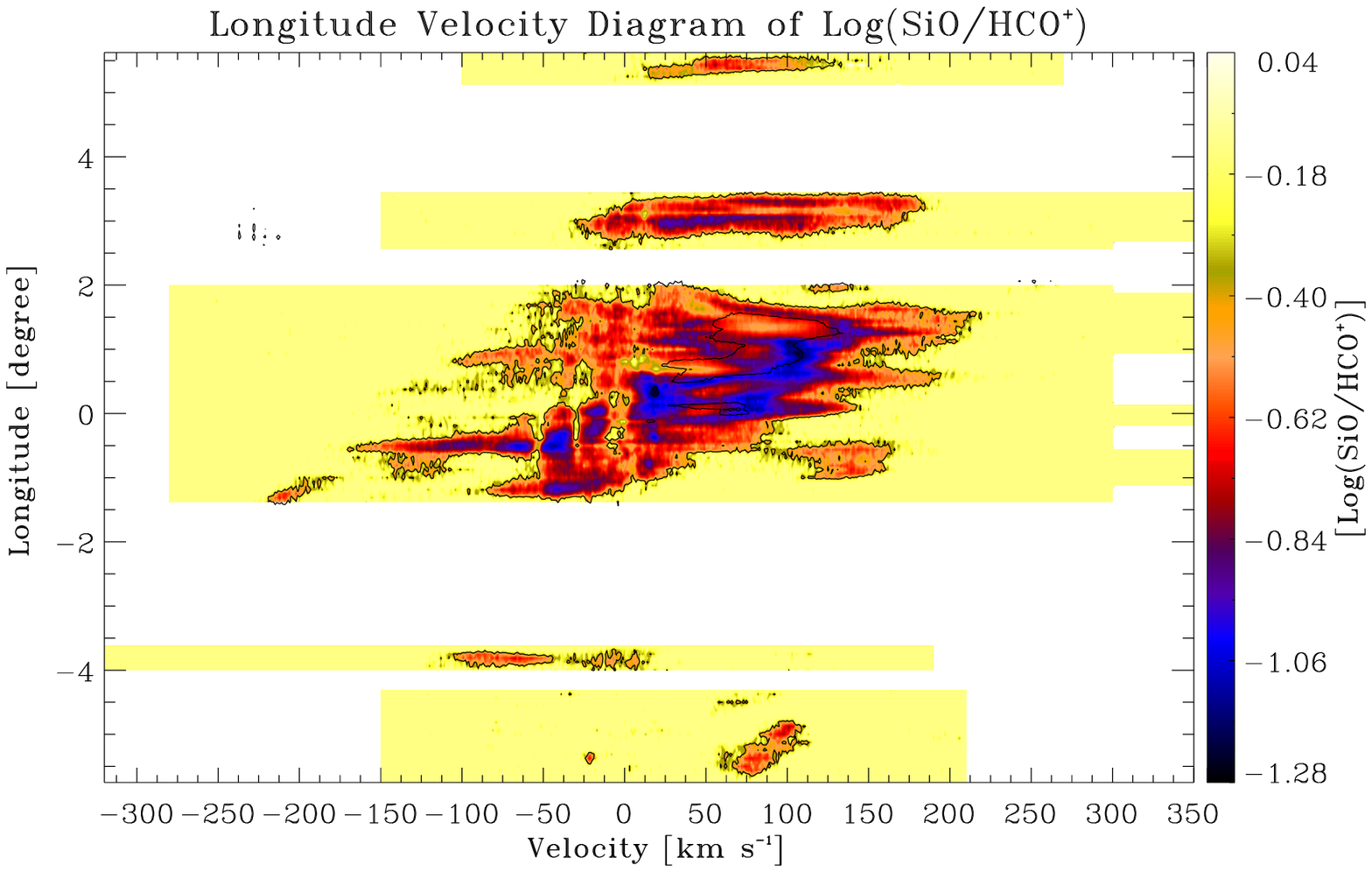}
\caption{Longitude-velocity emission comparison between the SiO and HCO+ emission. We plot
$\rm log(\int T(SiO (J=2\to1)) db/ \int T(HCO^+(J=1\to0)) db)$. In the region toward Sgr C, Sgr A, and Sgr D are
dominated by HCO+, and the region toward the 1.3 deg complex, Sgr B, M+3.2+0.3, and M+5.3-0.3 are dominated by
SiO.}
\label{comparacion_lv_SiO-HCO}
\end{center}
\end{figure*}
\indent To compare SiO and \HCO+ emission, we plot the logarithm of
the integrated intensity ratio for all positions where the emission of
both spectral lines is above $3 \sigma$ and in different velocity
channels. In regions where the emission is below this value we use the
$3 \sigma$ threshold.  We also compare the SiO emission with the
\H13CO+ emission, and use  the $3 \sigma$ threshold in both
transitions. \\
%
%Since \HCO+ has a much better signal-to-noise ratio that its rarer
%isotope \H13CO+, we will base out further analysis on this main
%isotope. 
\indent The \H13CO+ is useful since it is
optically thin and therefore traces the deeper regions of the
clouds. Because it has a high critical density of $\sim 10^{5}$ cm$^{-3}$
\citep{Tools2009}, it picks out the densest regions in our
maps. In Fig. \ref{comparacion_espacial_moleculas} we plot the
logarithm of the integrated intensity ratio for SiO and \HCO+ in the
velocity range from 30 to 130 \kms, and for SiO and \H13CO+ in the
velocity range from -110 to -50 \kms. In Appendix A, we plot the
logarithm of the integrated intensity ratio for velocity intervals of
50 \kms (Figs. A.3 and A.4). We can clearly identify
regions where the HCO$^{+}$ (blue regions), or where SiO dominates
(yellow and red regions). The SiO-dominated regions are, M$-$3.8+0.9 cloud (Fig. \ref{comparacion_espacial_moleculas}, and in
the velocity range from $v_{\rm LSR}= -100$ to $-50$ $\kms$ in
Fig. A.4), M+3.2+0.3 cloud, and M+5.3$-$0.3 cloud
(Fig. \ref{comparacion_espacial_moleculas}, and in the velocity range
from $v_{\rm LSR}=50$ to 150 \kms in
Fig. A.3), the 1.\deg3
complex ($\rm v>0 \kms$) and toward Sgr E region, both in negative
velocity and in forbidden velocity between $100<v<150$ $\kms$ (Fig. A.4). The \HCO+ is dominant toward
Sgr A ($-50<v<100$) and Sgr C ($-150<v<0$) in the CMZ. In the velocity
range of $v_{\rm LSR}= 0$ to $50$ \kms, we observe a very intense SiO
zone toward Sgr B, but this velocity range could be contaminated by
local gas seen by absorption in \HCO+, toward $v\sim 0$ $\kms$ (see
e.g. top of Fig. A.2), which could
increase the SiO to \HCO+ ratio emission (as can be seen in the
Fig. \ref{comparacion_lv_SiO-HCO} at $v\sim 0$ \kms.\\  
\indent In Fig. \ref{comparacion_lv_SiO-HCO}, we plot the logarithm of the
ratio of the intensities integrated in latitude between SiO and \HCO+
emission, using the $3 \sigma$ threshold. In the region toward Sgr C,
Sgr A and Sgr D are dominated by \HCO+, and the region toward the
1.3\deg complex,  Sgr B, M+3.2+0.3, and M+5.3$-$0.3  are dominated by
SiO.\\
%The high SiO to \HCO+
%intensity seen toward $l\sim 0\deg.75$ and $v\sim 0$ $\kms$ occurs due
%the \HCO+ absorption in the spiral arm, so it can not be interpreted
%as a SiO predominancy.
%
\indent To relate the observed line intensities and intensity
ratios to molecular column densities and abundance ratios, assumptions
on the excitation conditions of the gas are required. First of all, it
is necessary to estimate whether the observed transitions are
optically thick or optically thin.  In the case of \HCO+, we have
measurements of its rarer isotopomere \H13CO+. The $^{12}$C$/^{13}$C isotopic ratio in the Galactic
center region is about 20 \citep{Wilson_Matteucci_1992}. If both,
\HCO+ and \H13CO+ are optically thin in its $J=1\to0$ transitions one
would expect that their line intensity ratio is close to 20. On average, the measured line intensity ratio in the observed region is
typically between 10 and 30, with an average of 19.8 (see Fig. A.5). This indicates that the \HCO+ $(1-0)$
emission is indeed optically thin or just moderately optically thick
in most of the positions measured by us. This cannot be taken for
granted for other galactic centers; for example, in the nearby starbust
galaxy NGC253, the \HCO+ emission  is on average optically thick
\citep{Henkel_et_al_1993}.\\
\indent This allows column densities of the levels involved in
the transition to be determined \citep[see e.g.][for the corresponding
equations]{Mauersberger_Henkel_1991}. More difficult is the task of
determining the total column densities of the corresponding molecules
since, depending on the excitation conditions ($T_{\rm kin}$, $n({\rm
H}_2)$), the observed levels may represent only a small fraction of
the total column density. However, SiO and \HCO+ have
very similar dipole moments, namely 3.1 \citep{Raymonda_et_al_1970} and 3.9 Debye \citep{Botschwina_et_al_1993}, and
therefore their excitation conditions expressed in terms of critical
density should be similar.\\
\subsection{Intensity ratio of molecular emission}
\begin{figure*}
\hbox{
\includegraphics[width= 0.3 \textwidth, angle=90]{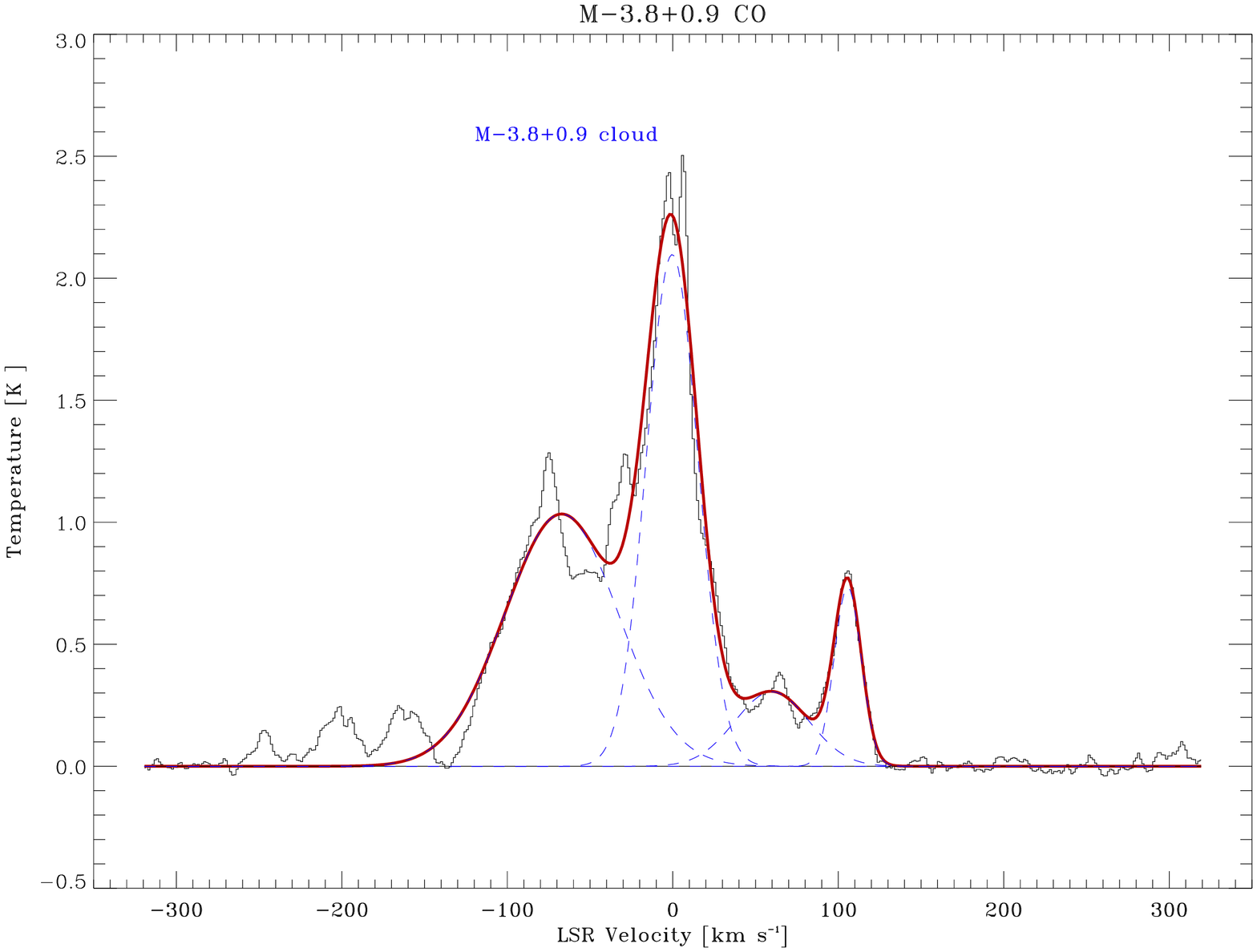}
\hspace{-0.3cm}
\includegraphics[width= 0.3 \textwidth, angle=90]{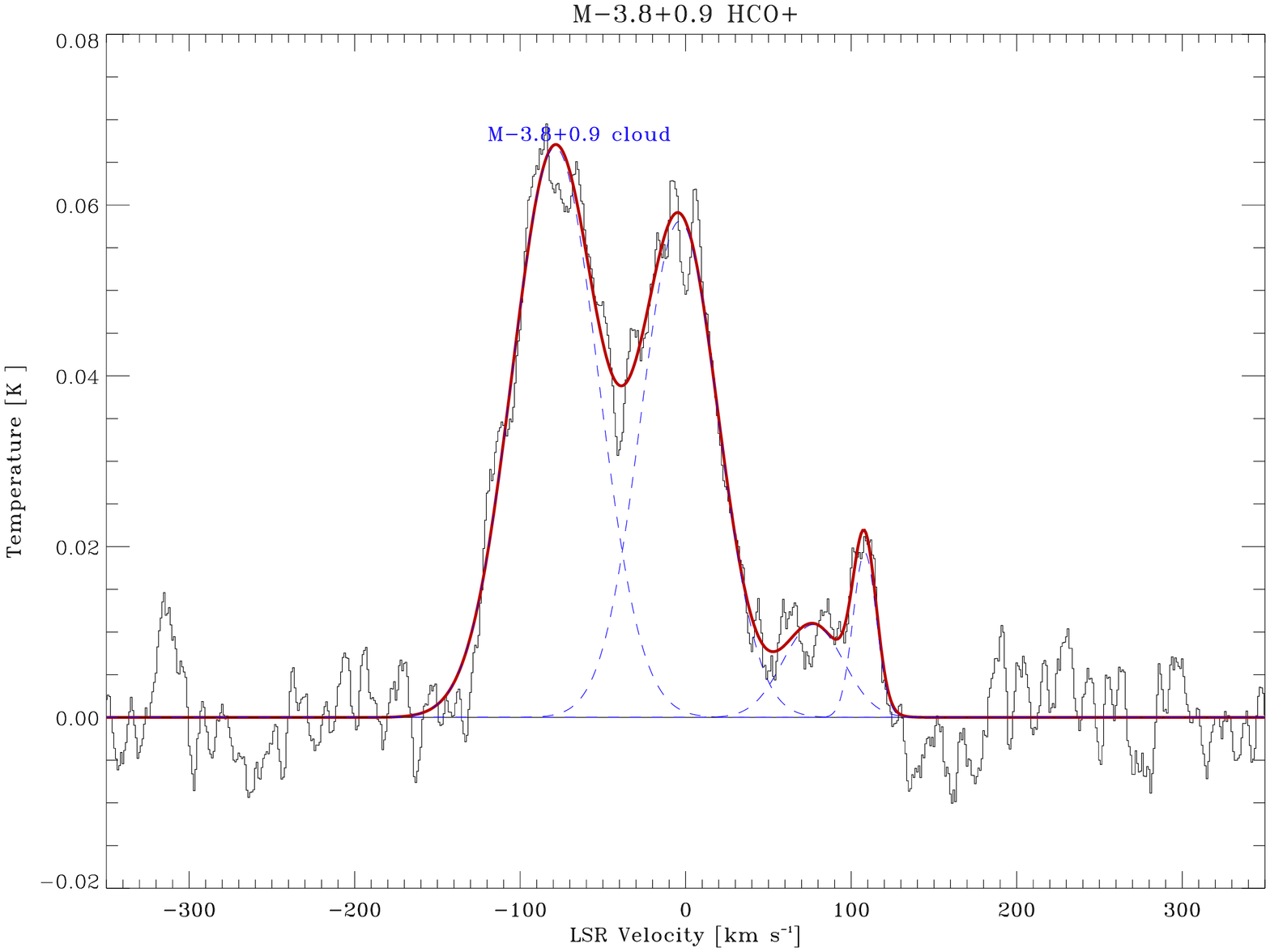}
}
\vspace{0.0cm}
\hbox{
\includegraphics[width= 0.3  \textwidth, angle=90]{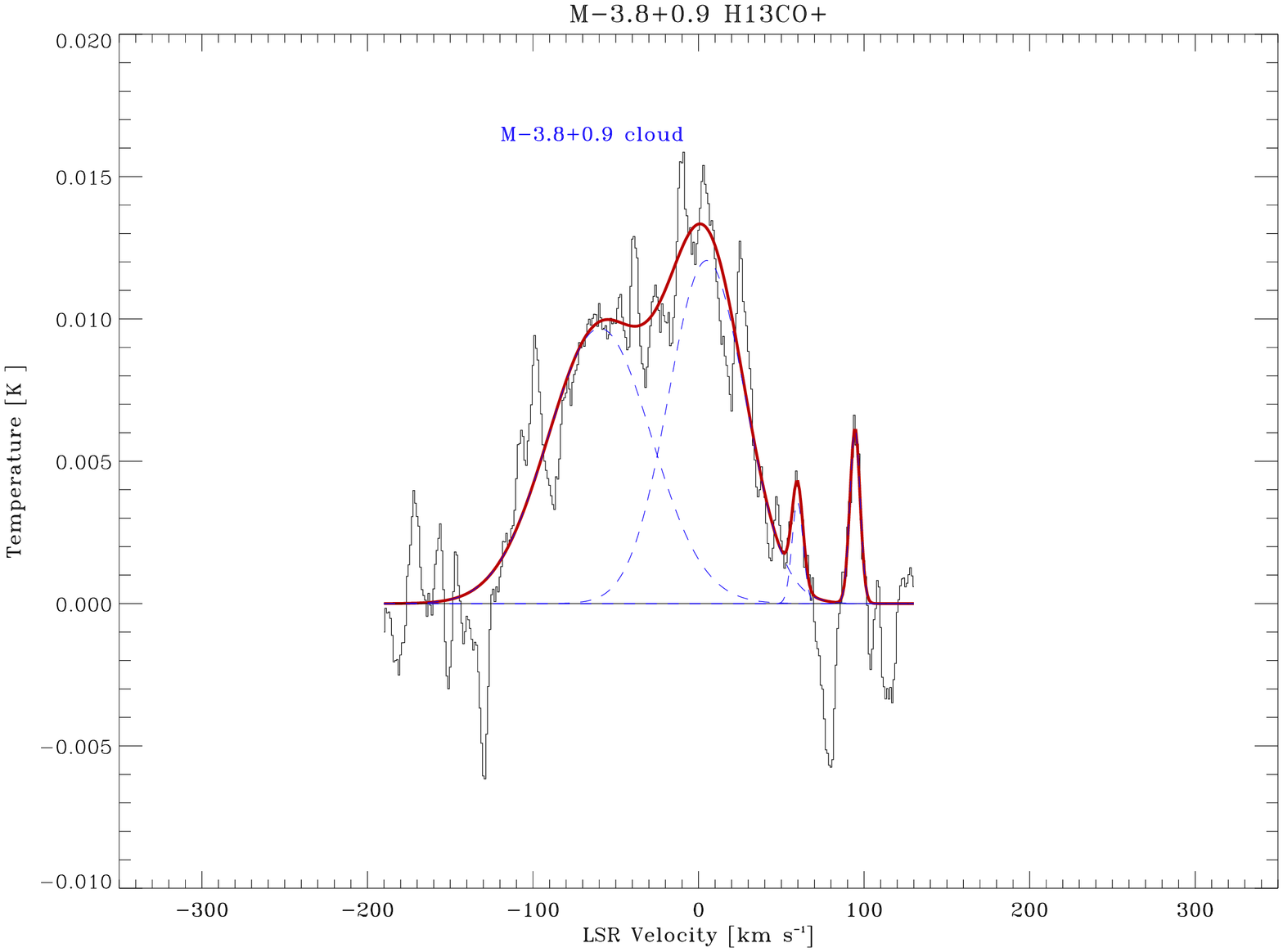}
\hspace{-0.3cm}
\includegraphics[width= 0.3 \textwidth, angle=90]{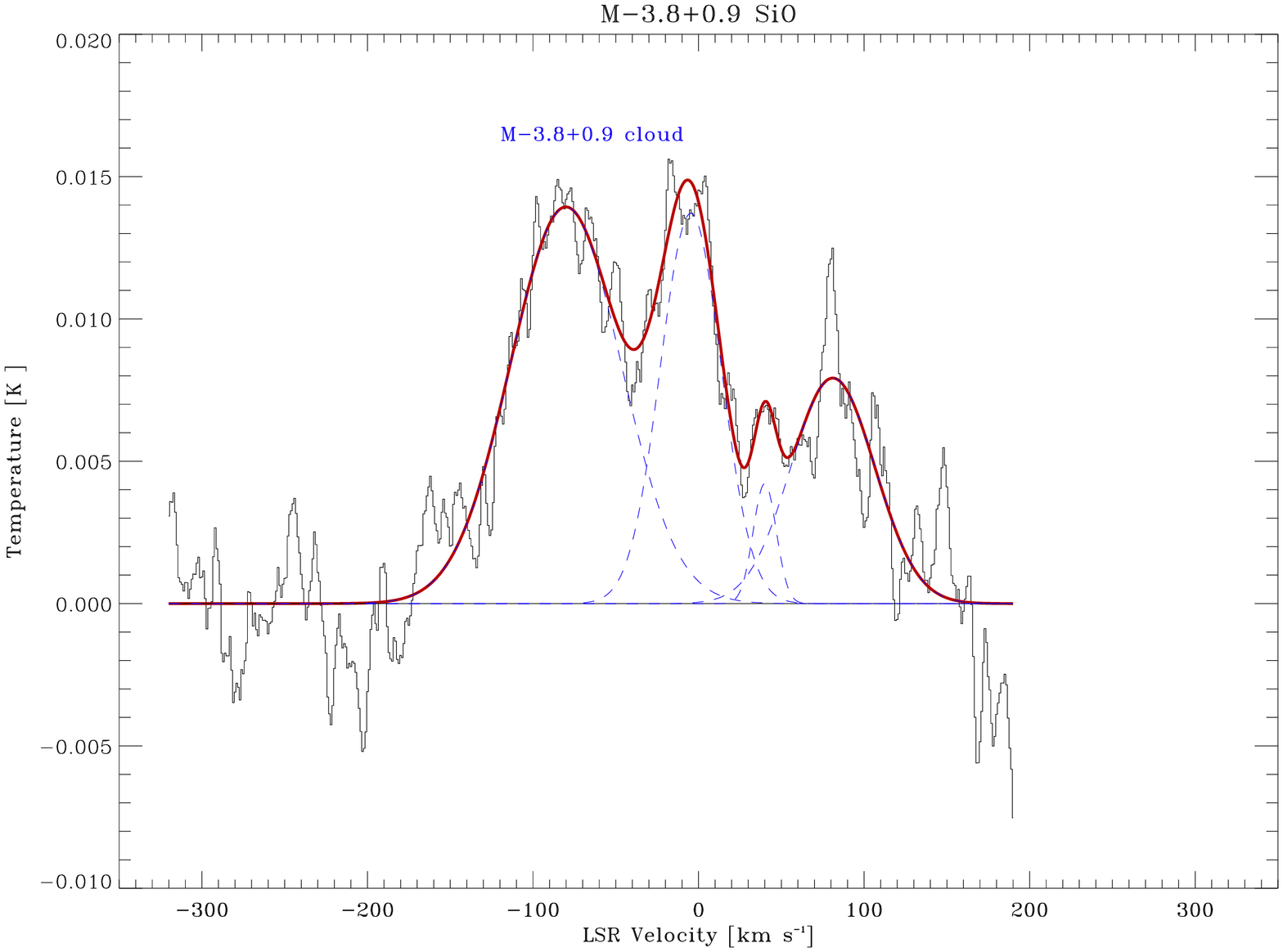}
}
\caption
{CO, HCO$^+$, SiO, and H$^{13}$CO$^+$ average spectra over the angular
size of M-3.8+0.9 cloud (from l= -4\deg.0 to -3\deg.625, and from
b=0\deg.5625 to 1\deg.1875). The angular size considered for each
region is listed in Table  2. The red lines indicate the Gaussian fit for the complete region and blue dashed lines show the Gaussian fits of each velocity components.}
\label{espectro_sumado_clumpC_paper}
\end{figure*}
\begin{figure*}
\begin{center}
\vbox{
\includegraphics[width= 0.8 \textwidth, angle=0]{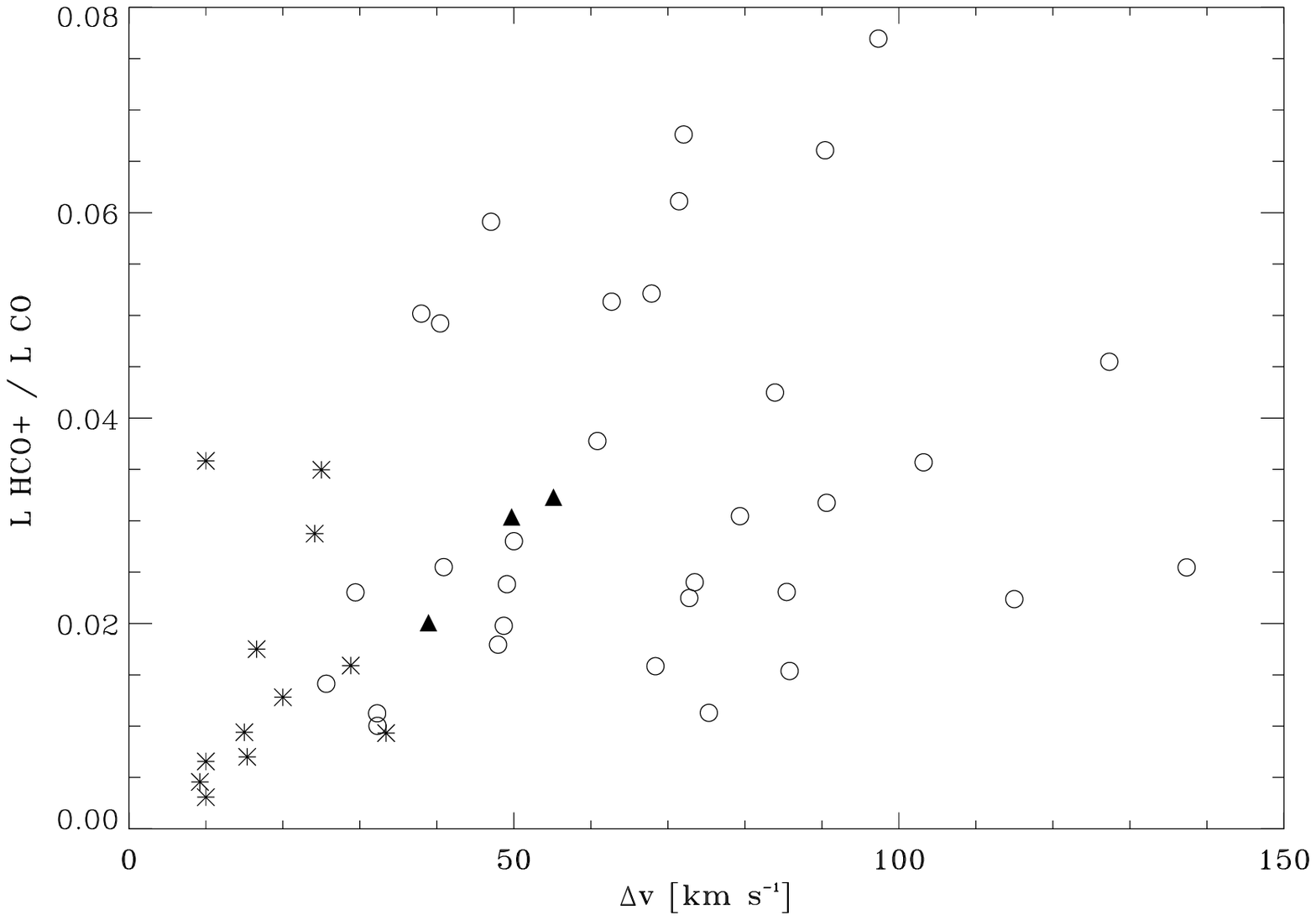}
\includegraphics[width= 0.8 \textwidth, angle=0]{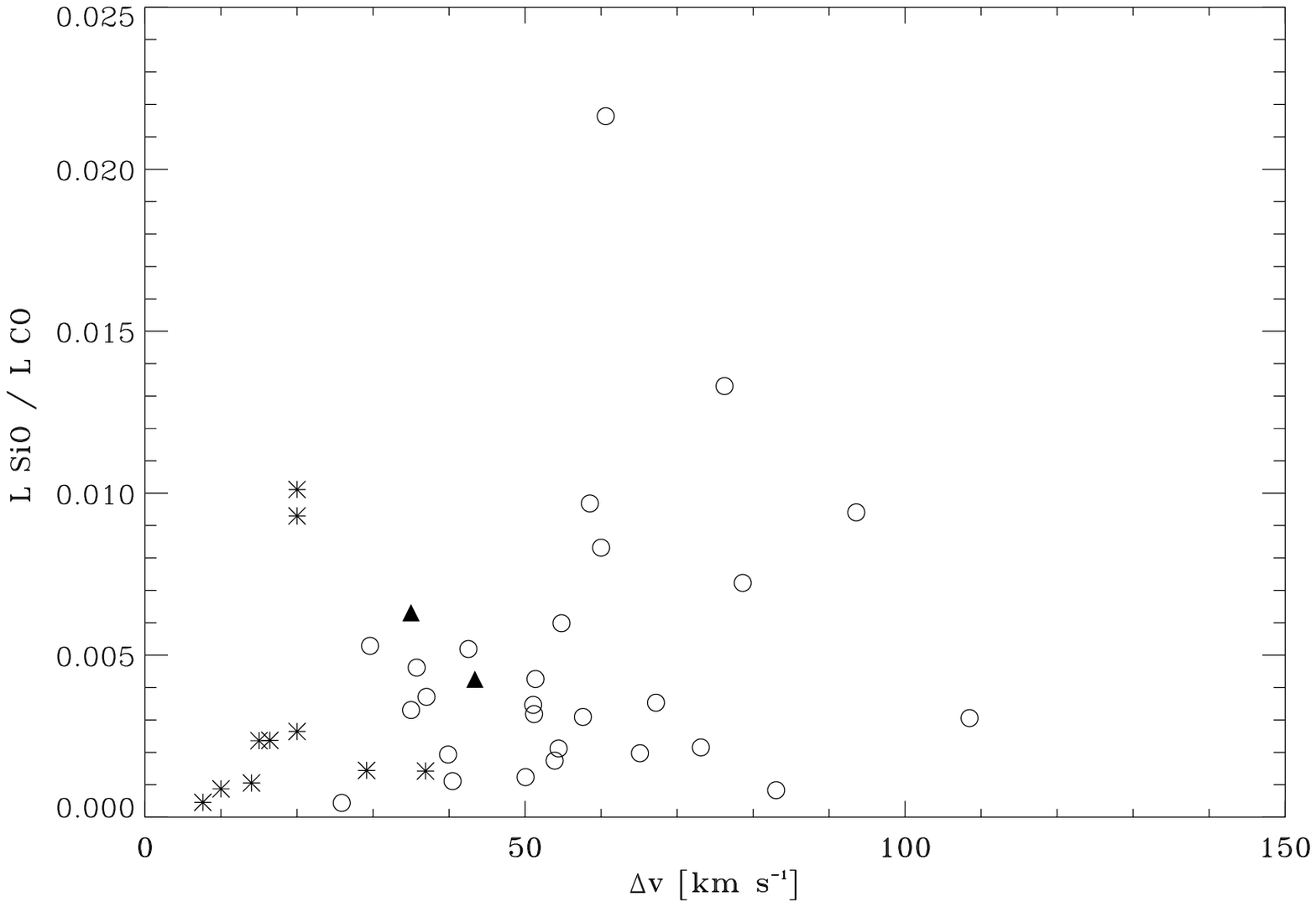}
}
\caption{Relationship between the luminosity ratio \HCO+ to CO (top)
  and the \HCO+ velocity width, and the luminosity ratio SiO to CO
  (bottom) and the SiO velocity width for each molecular cloud of the
  survey. Open circles denote Galactic center clouds, an asterisk the
  disk clouds, and filled triangles are clouds that probably are in an intermediate region, influenced by a bar, and that present large linewidth,
  probably due the strong Galactic center tidal forces in this
  region. There is a cloud (number 35 in SiO) with a large linewidth
  ($\sim 35$ \kms) in the bottom plot. This cloud is also not
  considered in our analysis owing the poor fit (see Fig. E.8).}
\label{luminosidad_survey}
\end{center}
\end{figure*}
\indent We also determined the ratio of the apparent luminosity S
between the molecular emissions of \HCO+ to CO, SiO to CO, and
\HCO+ to SiO, integrated over the observed regions to characterize
the physical properties of different clouds. The apparent luminosity S
is defined as the total emission integrated over velocity and solid
angle, in units of K \kms deg$^2$  \citep{Dame_et_al_1986}. The $\rm
HCO^+(J=1\to0)/CO(J=1\to0)$ ratio can be related to the ionization
fraction of the gas, and the intensity ratio $\rm
SiO(J=2\to1)/CO(J=1\to0)$ is  a measure of the amount of material
subject to shocks compared to ``quiescent'' gas. We also plot the
apparent luminosity ratio $\rm SiO(J=2\to1)/HCO^+(J=1\to0)$. To
define the molecular clouds, we use the average spectrum of each
region in \HCO+, SiO, \H13CO+, and CO from
\citet{Bitran_et_al_1997}. In the average spectrum, we perform
Gaussian fits to identify each molecular cloud. Different molecular
clouds can be distinguished by one dimensional Gaussian fits (Online
Appendix E), which yield temperature peaks (T$_o$), velocity
centers, and  velocity widths (FWHM) of the average spectra of the
different regions (Table \ref{ajuste_gausiano_nubes}). Also, we list
central positions for the clouds. All the values and errors in the
Table \ref{ajuste_gausiano_nubes} come from the Gaussian fits. Also,
we  assign locations to the clouds in this table. In cases where
the Gaussian fits did not give unique results or did not converge, the
values were obtained by visual inspection. Such cases are marked with
an asterisk.\\
\indent Clump-finding algorithms, such as
``Clumpfind'' \citep{Williams_et_al_1994}, have shown to be themselves useful for
identifying clumps in Galactic molecular cloud. In the present work, we
aim to identify the different velocity molecular clouds (with $10^5$ -
$10^6$ M$_o$) along the line of sight toward the Galactic center
region. It is not intended to derive the internal substructure within
every molecular cloud identified. That is why we only fit Gaussians in
the velocity dimension. We identified 51 molecular clouds, 33 of them
belonging to the Galactic center region and 18 to the
Galactic disk, local gas, or clouds along the line of sight. The
molecular clouds classified as outside the Galactic center are
characterized by narrow linewidths ($< 10$ \kms). However, there are
still some clouds classified as outside the Galactic center, which
present large linewidths. A possible reason is that the clouds could
be under the strong influence of the Galactic center tidal forces
(e.g. cloud numbers 44 and 48, see Table
\ref{ajuste_gausiano_nubes}). For example, cloud number 51 has a
large linewidth in \HCO+. This cloud belongs to the 135 \kms arm
\citep{Bania_1980}, which is supposedly located outside from the
Galactic center, but it is strongly influenced by it.\\
\indent The apparent luminosity for each molecular component was obtained using
\begin{eqnarray}
\label{luminosidad}
S&=& \int_{\Omega_s} \int_v T^*_{\rm A} \rm dv \rm d\Omega , %, \nonumber\\
%&=& \Delta l \Delta b \int_v \sum_{positions}T^*_{\rm A} \rm dv  \rm{\quad [K \ km\, s^{-1}(degree)^2 ] ,}
\end{eqnarray}
where $T^*_{\rm A}$ is the antenna
temperature. Figure \ref{espectro_sumado_clumpC_paper} shows an example
of the average spectra and the Gaussian fit. In this cloud the different
intensities of the emission in the main velocity component ($v\sim -79$
\kms) belonging to the Galactic center are evident. The HCO$^+$ and SiO emission
in the main velocity component have a noticeable increase of the
intensity when compared with, e.g., the gas at velocities $v\sim0$
\kms, which, presumably correspond to gas in the line of sight\footnote{The line of sight components are shown in the Appendix
A, Fig. A2.2 as a narrow emission ($\sim 5-10$ \kms), whereas the
Galactic center emission is characterized by broad velocity width
lines ($\gtrsim 50$ \kms). Thus, the emission coming from the region
at $l\sim -0\deg.4$ to $l\sim 0\deg.6$ and $b\sim -0\deg.4375$ to
$b\sim -0\deg.375$ at $\sim 0$\kms  corresponds to local gas.}. In the
CO emission, the main component ($v\sim -79$ \kms) show less emission
when compared with the gas in the line of sight at $v\sim0$ \kms. This
plot clearly shows the differences in the molecular gas in
the Galactic center and in the disk. In the Appendix D, we show the
Gaussian fits for all the molecular complexes.\\
\indent Figures A.6 and A.7 show the ratio of
\HCO+ and SiO to CO luminosities, respectively, while
Fig. A.8 shows the ratio of SiO to \HCO+ luminosities for
each molecular cloud.  The
``main component'' is the most prominent Galactic center cloud in the
region (see Table \ref{main_component}), and it was identified by
\citet{Bitran_1987}.
%The clouds belonging to the Galactic center show a
%higher \HCO+ to CO luminosity ratio than for the disk.
It is noticeable that we could identify some SiO clouds as belonging
to the local gas and/or spiral arms, while it is supposed that SiO
only traces the gas belonging to the Galactic center. From their
velocities and line shape these clouds appear to be in the Galactic
disk rather than in the Galactic center. That they are
emitting SiO radiation would, however, indicate a location within the
Galactic center region. A more detailed study of these clouds would be
interesting, since they are either Galactic center clouds with an
unusual velocity footprint or they are disk clouds with unusual
chemistry and/or excitation conditions.
%, which could mean
%that the ionization fraction in
%the molecular clouds in the Galactic center is lower than in disk
%clouds so the ion-slip heating is more effective in that region. 
The average of the ratio of \HCO+ to CO luminosity in clouds belonging
to the Galactic center is $0.035 \pm 0.003$ and for disk clouds is $
0.015 \pm 0.004$. The higher intensity ratios are found toward cloud 9
in Sgr B, cloud 4 in Sgr A, cloud 17 in Sgr D, cloud 23 in Sgr E, and
cloud 25 in the 1\deg.3 complex.  In the same way, we display the
ratio between SiO and CO. The average of the ratio of SiO to CO
luminosity in clouds belonging to the Galactic center is $0.0049 \pm
0.0005$ and for disk cloud is $0.0034 \pm 0.0009$. A higher abundances
of $\rm SiO(J=2\to1)/CO(J=1\to0)$ is observed in the M+3.2+0.3 cloud,
1\deg.3 complex, and in the M+5.3$-$0.3 cloud. The luminosity ratio of
$\rm SiO(J=2\to1)/HCO^+(J=1\to0)$ in Fig. A.8, gives an
average of $0.15\pm 0.002$ for the Galactic center and $0.26\pm 0.05$
for the disk clouds. The higher ratios in the Galactic center are
found in the M+3.2+0.3 cloud, M+5.3$-$0.3 cloud, and Sgr D region, and
the lower in Sgr A, Sgr C, and Sgr B.  For the clouds belonging
to the Galactic disk, the average was obtained without considering the
clouds with large linewidths discussed before (clouds number 44, 48, and
51), and for the cloud belonging to the Galactic center we did not
consider the clouds that present self absorption in \HCO+ and CO,
which would decrease the integrated intensity of the cloud (clouds
number 2, 11, and 21).\\
\setcounter{table}{3}
\begin{table*}
\caption [] {Velocity components of each region with the longitude and latitude ranges used to defined different regions.}
\label{main_component}
\centering 
\begin{tabular}{c c c c c c c c c}  
  \hline\hline                        % inserts double horizontal lines
   Zone& Region & Longitude & Latitude &\multicolumn{5}{c}{Velocity center [\kms]} \\
       &        &  [deg]    &  [deg]   &  v$_{\rm LSR}$ &  v$_{\rm LSR}$&  v$_{\rm LSR}$ &  v$_{\rm LSR}$& v$_{\rm LSR}$\\\hline\hline  
   CMZ & Sgr A           &$-0.3125<l< 0.3125$&$-0.5 <b< 0.5$&$-$76    & $-$28$^{\mathrm{a}}$    & 17    &{\bf 61}& 135 \\\hline
   CMZ & Sgr B           &$ 0.375<l<0.8125$&$ -0.5<b<0.5$ & $-$56    &{\bf 42}                & 95    & 144   &     \\\hline
   CMZ & Sgr C           &$-0.6875 <l<-0.375 $&$-0.5 <b<0.5$& $-$103 &{\bf$-$43}$^{\mathrm{b}}$& 22    & 132   &     \\\hline
   CMZ & Sgr D           &$ 0.875<l< 1.1875$&$ -0.5625<b<0.5625$& $-$56    & $-$10              & 27    &{\bf 88}& 180 \\\hline
   CMZ & Sgr E           &$ -0.75<l<-1.5 $&$ -0.5<b<0.5$& {\bf $-$203}   & $-$124    &$-$31     & 17& 131 \\\hline

   CMZ & 1\deg.3 complex &$1.25 <l< 2.0$&$-0.5625 <b< 0.5625$&$-$22    &{\bf 83}               & 178   &       &     \\\hline
   PMZ & M+3.2+0.3       &$ 2.5625<l<3.5 $&$-0.25 <b<0.875$& 32     &{\bf 103}               & 159   & 219   &     \\\hline
   PMZ & M$-$5.3+0.4       &$ -5.75<l<-4.75$&$-0.125 <b<0.5625$&$-$78  &$-$43  & $-$21 &5          &    {\bf 86}\\\hline
   PMZ & M$-$4.4+0.6       &$ -4.6875<l<-4.3125$&$ 0.4375<b<0.8125$& $-$70  & $-$45   &$-$30         &{\bf 71}&            \\\hline
   PMZ & M$-$3.8+0.9       &$ -4.0<l<-3.6875$ &$0.5625 <b<1.1875$&{\bf $-$79}& $-$4                & 77    & 108   &     \\\hline
   PMZ & M+5.3$-$0.3       &$ 5.125<l<5.5625$ &$ -0.6875<b< 0.125$& $-$28    & 23                & 59    &{\bf 98}&150 \\\hline
\end{tabular}
\begin{list}{}{}
\item[] The velocity component are defined by HCO$^+$ Gaussian fits, see Appendix E. The main cloud is indicated in bold face.
\item[$^{\mathrm{a}}$]  has 2 velocity components.
\item[$^{\mathrm{b}}$]  has 3 velocity components.
\end{list}
\end{table*} 
\indent We also investigated the relationship between the $\rm HCO^+(J=1\to0)/CO(J=1\to0)$ and
$\rm SiO(J=2\to1)/CO(J=1\to0)$ luminosity ratio and the velocity width of the respective clouds in
Fig. \ref{luminosidad_survey}. Here we show disk clouds, clouds in the Galactic center, and cloud that presumably
  belong to the Galactic disk but they present large linewidth,
  probably because of the strong Galactic center tidal forces in this
  region. It is evident, in general, that
Galactic center clouds show higher $\rm HCO^+(J=1\to0)/CO(J=1\to0)$ and $\rm SiO(J=2\to1)/CO(J=1\to0)$
luminosity ratios and larger linewidths than disk clouds. 
\subsection{Comparison with previous work}
\indent As shown before, we can distinguish regions where either SiO
or \HCO+ dominates. Roughly, in the CMZ at longitudes lower than
$l\sim 0\deg.6 $, \HCO+ dominates, and at longitudes $l>0\deg.6$, SiO
prevails, indicating shock. Nevertheless, we find clouds with
an enhancement of SiO toward lower longitudes in the CMZ. For cloud 4
in Sgr A and cloud 7 in Sgr B, the SiO is also intense. In the PMZ, the
clouds M+3.2+0.3, M+5.3$-$0.3, and M$-$3.8+0.9 show an enhancement of
SiO, which is a clear signal of shocks.\\
\indent The SiO abundance can be increased, e.g., as a consequence of
cloud-cloud collisions, interactions with supernova  remnants,
expanding bubbles, and large-scale dynamics in the Galactic center. The
SiO predominance that we find in clouds 4 and 7 has been
noted by other authors. \citet{Martin-Pintado_et_al_1997} show that
SiO emission is detected throughout the whole Galactic center
region. They related the intense SiO
emission that they found toward the Sgr A molecular complex
(M$-$0.13$-$0.08 which is the 20 $\kms$ cloud, M$-$0.02$-$0.07 which
correspond to the 50 $\kms$ cloud, and a condensation close to Sgr A*)
to the interaction of the molecular clouds with nearby supernova
remnants. Their SiO emission spots could correspond to our cloud
number 4 in Sgr A region, but in our data they are blended because of
our lower resolution (see Table
\ref{main_component}). \citet{Minh_et_al_1992} also found high
abundances of SiO and \HCO+ toward Sgr A region, which indicate that
shock chemistry and ion-molecule reactions are important in this
region.\\
\indent The enhancement of SiO
that we found toward greater longitudes ($l>0\deg.6$ and in the PMZ)
has been also reported by \citet{Huettemeister_et_al_1998}.  They
performed multiline observations of the C$^{18}$O and also SiO
isotopes in the Galactic center region toward 33 selected positions
from the CS survey of \citet{Bally_et_al_1987}.
%The observed region includes the CMZ,
%M+3.2+0.3  cloud, and a cloud belonging to the M$-$5.3+0.4 cloud (clump 1). 
All the sources were easily detected in SiO, where the higher abundances are
found at $l>0\deg.8$. They found two regimens of densities and temperatures, one dense
and cool, and other thin and hot, which are in pressure equilibrium,
where the SiO emission arise in a cool, moderately dense component
\citep{Huettemeister_et_al_1998}.
%However, one source in their sample, M+1.31$-$0.13, with a large SiO
%abundance, present a consistent solution of temperature and densities where the SiO emission comes from a thin, hot gas,
%which suggests that the SiO, as in the disk, is formed in shocks}. 
%Otherwise, in the Galactic disk the dense cores are hot, and the SiO emission arise from a hot dense gas.  
The enhancement of the SiO emission was related to the large-scale gas
dynamics in the Galactic center region where the movement of the gas
can be understood as the response of a rapidly 
rotating bar potential \citep{Binney_et_al_1991}, and the higher
abundances of SiO can be identified with the collision region.  This
molecular cloud has also been studied by \citet{Tanaka_et_al_2007}. They
identified 9 expanding shells with broad-velocity-width features in their HCN
and \HCO+ maps and isolated SiO clouds that should be related to the
expanding shells. They propose that the expanding shells may be in the
early stage of supperbubble formation caused by massive cluster
formation or continuous star formation $10^{6.8-7.6}$ years ago.
Both \citet{Huettemeister_et_al_1998} and
\citet{Martin-Pintado_et_al_1997} observed a decrease of $X(\rm SiO)$
in the CMZ (between Sgr B2 and Sgr C, $-0.\deg35$$<l<0.\deg6$) with
respect to higher longitudes ($l>0.9$), which is also seen in our
data. Figures A.6 and A.7 show that the SiO
emission mainly comes from $l>0\deg.6$ and that \HCO+ emission is dominant in this region, which shows the densest zones where star formation is ongoing.\\
\indent In this work we relate the SiO enhancement
throughout the Galactic center region to the Giant molecular loops scenario proposed by \citet{Fukui_et_al_2006}. \citet{Fukui_et_al_2006} observed an area of 240 square degrees 
toward  \mbox{$-12$\deg $<l<$ $12$\deg} and \mbox{$-5$\deg $<b<$ $5$\deg} in
\mbox{$^{12}$CO(1-0)} using the NANTEN \mbox{$4$ m} telescope from Nagoya
University (the NANTEN Galactic plane survey, GPS). They find huge structures in loop shapes, and propose 
that there are ``giant molecular loops'' (huge loops of dense
molecular gas with strong velocity dispersions) at the Galactic center, formed by a 
magnetic buoyancy caused by the Parker instability. The loops have two
``foot points'', one at each end, which are produced when the
gas inside the loops flows down to the disk by stellar gravity and forms
shock fronts above the disk. This scenario is supported by numerical
simulations (\citealp{Matsumoto_et_al_1988,Machida_et_al_2009,
  Takahashi_et_al_2009}) and by the broad velocity features of \mbox{$\sim$ 40} to \mbox{80 \kms} observed 
by \citet{Fukui_et_al_2006}. The shocked regions detected in SiO in
the present work are correlated with the foot points they found. The enhancement of SiO
emission, in comparison with the \HCO+ emission that we found in the
M$-$3.8+0.9 cloud (Fig. A.3, A.4), is correlated with the foot
point of the loop 1 (toward $l\sim -4\deg$ to $-2\deg$, in the velocity
range from $-180$ to $-90 \kms$) and loop 2 (toward $l\sim -5\deg$ to $-4\deg$, in the
velocity range of $-90$ to $-40 \kms$). Those features are studied in detail by \citet{Torii_et_al_2009a,Torii_et_al_2009b}.
The enhancement of SiO in  M+3.2+0.3 and M+5.3$-$0.3 clouds are 
correlated with the foot point of the loop at 
positive longitudes, shown in the Fig. S6 on the ``Supporting
Online Material'' in \citet{Fukui_et_al_2006}. This feature is placed at positive longitudes between $l\sim 3\deg$ to $5\deg$. The  enhancement of SiO found toward
$l\sim -1\deg$ corresponds to the location of loop 3, which has recently been discovered by \citet{Fujishita_et_al_2009}. This loop is located toward $l\sim -5\deg$ to $-1\deg$ in the velocity range of $20$ to $200 \kms$ \citep{Fujishita_et_al_2009}. The coincidence of the enhancement of SiO to the \HCO+ in the ``foot point'', together with the high-velocity width of the clouds belonging to the Galactic center  (see Table \ref{ajuste_gausiano_nubes}), support Fukui's
scenario. This association will be addressed in more detail in a subsequent paper. 
%
%______________________________________________________________
%
\section{Conclusions}
\begin{enumerate}
\item All of the species measured in this work, \HCO+, SiO, and \H13CO+, have
been detected throughout the Galactic center region. We find the
characteristic asymmetry in longitude found for many other species, with most
of the emission toward $l>0$ and v$>0$. We identify 51 molecular
clouds, where 33 belong to the Galactic center region and 18 to the Galactic disk or local gas.  

\item The luminosity ratios $\rm SiO(J=2\to1)/CO(J=1\to0)$ and $\rm HCO^+(J=1\to0)/CO(J=1\to0)$, as well as the velocity
widths, are higher for Galactic center clouds than for typical disk clouds. The
highest $\rm SiO(J=2\to1)/CO(J=1\to0)$ luminosity ratios for the 
Galactic center region correspond, in general, to the highest velocity
widths. The average of the luminosity ratio of  $\rm SiO(J=2\to1)/CO(J=1\to0)$  in clouds
belonging to the Galactic center region is $0.0049 \pm 0.0005$ and for disk
clouds is $0.0034 \pm 0.0009$. The luminosity ratio of $\rm HCO^+(J=1\to0)/CO(J=1\to0)$ in the
Calactic center is $0.035 \pm 0.003$, and for disk clouds is $ 0.015\pm 0.004$.

\item The clouds M+3.2+0.3, M-3.8+0.9, M+5.3-0.3, and 1.3\deg
complex show high SiO to \HCO+ ratios, which may indicate the
importance of shocks as heating sources. Toward the densest regions,
the $\rm SiO(J=2\to1)/HCO^+(J=1\to0)$ ratio is low (Sgr A and Sgr B regions). 

\item The SiO emission can be correlated with several phenomena. The SiO
predominance over the \HCO+ emission could be related to the molecular
loops, formed by a Parker instability, where the shocks are ongoing.
\end{enumerate}

%%%%%%%%%%%%%%%%%%%%%%%%%%%%%%%%%%%%%%%%%%%%%%%%%%%%%%%%%%%%%%%%%%%%%%%%%%%%%%%%%%%%%%%
\begin{acknowledgements}
      We acknowledge support by the Chilean Center for Astrophysics
      FONDAP N 15010003 and by Center of Excellence in Astrophysics
      and Associated Technologies PFB 06. D.R. and R.M. were supported
      by DGI grant AYA 2008-06181-C02-02. We thank Fernando Olmos for
      help with the observations. We are grateful to the personnel and
      students from Nagoya University who supported observations at
      the telescope and keep the data reduction pack at Cerro
      Cal\'an. We also want to thanks Jes\'us Mart\'{i}n-Pintado for
      helpful discussions. We thank the referee, Y. Fukui, and the
      editor of A\&A, M. Walmsley, for valuable comments.

\end{acknowledgements}

\bibliographystyle{aa} % style aa.bst
\bibliography{referencias} % your references Yourfile.bib
\setcounter{table}{2}
{\small
\longtab{3}{
\begin{longtable}{llllllll}
\caption{\label{ajuste_gausiano_nubes} Gaussian fits of the each component of the  molecular clouds}\\
\hline\hline
Region & Cloud & Line &  Central Velocity & Velocity Width$^{\mathrm{a}}$ & T$_0$ & Luminosity & Associated\\
       &  Number &      & [km s$^{-1}$]  &   [km s$^{-1}$]   &[K]  & [K km s$^{-1}$degree$^2$]&Object\\\hline 
\endfirsthead
\caption{continued.}\\
\hline\hline
Region & Cloud & Line &  Central Velocity & Velocity Width & T$_0$ & Luminosity & Associated\\
       &  Number &      & [km s$^{-1}$]  &   [km s$^{-1}$]   &[K]  & [K km s$^{-1}$degree$^2$]&Object\\\hline 
\hline
\endhead
\hline
\endfoot
Sgr A & 1 &CO & -126   $\pm$ 8 & 60  $\pm$ 18 & 1.3    $\pm$ 0.2    &   56.5 $\pm$   19.6 & EMR \\\hline
Sgr A & 1 &HCO$^+$ & -76  $\pm$ 2 & 127 $\pm$ 3  & 0.032  &    2.57 $\pm$    0.10 & EMR \\\hline
Sgr A & 1 &SiO & -110  $\pm$ 3 & 108 $\pm$ 8  & 0.0045 $\pm$ 0.0003 &    0.17 $\pm$    0.02 & EMR \\\hline
Sgr A & 2 &CO & -34    $\pm$ 18 & 68 $\pm$ 42 & 1.9    $\pm$ 0.3    &   95.7 $\pm$   61.0 &   \\\hline
Sgr A & 2 &HCO$^+$ & -39  $\pm$ 1  & 12 $\pm$ 1  & 0.066    &    0.5  &   \\\hline
Sgr A & 2 &HCO$^+$ & -18  $\pm$ 1  & 13 $\pm$ 1  & 0.10    &    0.85  &   \\\hline
Sgr A & 2 &H$^{13}$CO$^+$ & -26 $\pm$ 10 & 74$\pm$ 11 & 0.004  $\pm$ 0.002  &    0.12 $\pm$    0.05 &   \\\hline
Sgr A & 3 &CO & 11     $\pm$ 5  & 36 $\pm$ 14 & 2.5    $\pm$ 1.6    &   65.7 $\pm$   50.9 & MM \\\hline
Sgr A & 3 &HCO$^+$ & 17   $\pm$ 1  & 24 $\pm$ 1  & 0.12    &    1.9  & MM \\\hline
Sgr A & 4 &CO & 66     $\pm$ 8  & 71 $\pm$ 38 & 2.3    $\pm$ 0.2    &  124 $\pm$   67 & Sgr A cloud \\\hline
Sgr A & 4 &HCO$^+$ & 61   $\pm$ 1  & 72 $\pm$ 1  & 0.19    &    8.4  & Sgr A cloud \\\hline
Sgr A & 4 &SiO & 44    $\pm$ 1  & 94 $\pm$ 1  & 0.035  &    1.16  & Sgr A cloud \\\hline
Sgr A & 4 &H$^{13}$CO$^+$ & 40 $\pm$ 5  & 96 $\pm$ 6  & 0.016 $\pm$ 0.001 &    0.55 $\pm$    0.05 & Sgr A cloud \\\hline
Sgr A & 5 &CO & 147    $\pm$ 31 & 72 $\pm$ 49 & 0.8    $\pm$ 0.3    &   44 $\pm$   35 & EMR \\\hline
Sgr A & 5 &HCO$^+$  & 135 $\pm$ 4 & 79 $\pm$ 5 & 0.027 $\pm$ 0.001 &    1.3 $\pm$    0.1 & EMR \\\hline
Sgr B & 6 &CO    & -61 $\pm$ 7 & 100 $\pm$ 15 & 1.8 $\pm$ 0.1 &  110 $\pm$   17 & EMR \\\hline
Sgr B & 6 &HCO$^+$  & -56 $\pm$ 1 & 115 $\pm$ 3 & 0.046  &    2.47  & EMR \\\hline
Sgr B & 6 &SiO & -49 $\pm$ 3 & 83 $\pm$ 7 & 0.0053  &    0.092  & EMR \\\hline
Sgr B & 7* &CO & 30 $\pm$ 5 & 60 $\pm$ 5 & 4.1 $\pm$ 0.1 &  147 $\pm$   12 & Sgr B cloud \\\hline
Sgr B & 7 &HCO$^+$ & 42 $\pm$ 1 & 63 $\pm$ 1 & 0.26  &    7.56  & Sgr B cloud \\\hline
Sgr B & 7 &SiO & 38 $\pm$ 1 & 59 $\pm$ 1 & 0.12  &    1.43  & Sgr B cloud \\\hline
Sgr B & 7 &H$^{13}$CO$^+$ & 43 $\pm$ 1 & 55 $\pm$ 1 & 0.044  &    0.505  & Sgr B cloud \\\hline
Sgr B & 8* &CO & 90 $\pm$ 5 & 35 $\pm$ 5 & 2.75 $\pm$ 0.05 &   58 $\pm$    8 &   \\\hline
Sgr B & 8 &HCO$^+$ & 95 $\pm$ 1 & 47 $\pm$ 1 & 0.16  &    3.4  &   \\\hline
Sgr B & 8 &SiO & 93 $\pm$ 1 & 37 $\pm$ 1 & 0.028  &    0.214  &   \\\hline
Sgr B & 8 &H$^{13}$CO$^+$ & 94 $\pm$ 1 & 26 $\pm$ 1 & 0.013  &    0.070  &   \\\hline
Sgr B & 9 &CO & 170 $\pm$ 12 & 56 $\pm$ 19 & 1.0 $\pm$ 0.3 &   33 $\pm$   15 &   \\\hline
Sgr B & 9 &HCO$^+$ & 144 $\pm$ 3 & 97 $\pm$ 4 & 0.056  &    2.6  &   \\\hline
Sgr C & 10 &CO & -123 $\pm$ 5 & 80 $\pm$ 14 & 1.8 $\pm$ 0.2 &   88 $\pm$   18 & EMR \\\hline
Sgr C & 10 &HCO$^+$ & -103 $\pm$ 1 & 91 $\pm$ 1 & 0.095 &    2.8  & EMR \\\hline
Sgr C & 10 &SiO & -116 $\pm$ 1 & 50 $\pm$ 2 & 0.012  &    0.109 & EMR \\\hline
Sgr C & 10 &H$^{13}$CO$^+$ & -99 $\pm$ 1& 29 $\pm$ 1 & 0.006  &    0.031  & EMR \\\hline
Sgr C & 11 &CO & -64 $\pm$ 4 & 16 $\pm$ 11 & 1.1 $\pm$ 0.5 &   11 $\pm$    9 & Sgr C cloud \\\hline
Sgr C & 11 &CO & -41 $\pm$ 3 & 15 $\pm$ 8 & 1.6 $\pm$ 0.7 &   14 $\pm$    10 & Sgr C cloud \\\hline
Sgr C & 11 &CO & -24 $\pm$ 3 & 6 $\pm$ 7 & 0.9 $\pm$ 0.8 &    3 $\pm$    4 & Sgr C cloud \\\hline
Sgr C & 11 &HCO$^+$ & -65 $\pm$ 1 & 15 $\pm$ 1 & 0.079  &    0.38  & Sgr C cloud \\\hline
Sgr C & 11 &HCO$^+$ & -41 $\pm$ 1 & 15 $\pm$ 1 & 0.16  &    0.79  & Sgr C cloud \\\hline
Sgr C & 11 &HCO$^+$ & -21 $\pm$ 1 & 11 $\pm$ 1 & 0.067  &    0.235  & Sgr C cloud \\\hline
Sgr C & 11 &SiO & -50 $\pm$ 1 & 43 $\pm$ 2 & 0.019 $\pm$ 0.002 &    0.15 $\pm$    0.02 & Sgr C cloud \\\hline
Sgr C & 11 &H$^{13}$CO$^+$ & -54 $\pm$ 1 & 33 $\pm$ 2 & 0.011 $\pm$ 0.001 &    0.064 $\pm$    0.007 & Sgr C cloud \\\hline
Sgr C & 12* &CO & 10 $\pm$ 5 & 40 $\pm$ 5 & 4.1  &   99 $\pm$   12 &   \\\hline
Sgr C & 12 &HCO$^+$ & 22 $\pm$ 1 & 73 $\pm$ 1 & 0.099  &    2.37  &   \\\hline
Sgr C & 12 &SiO & 3 $\pm$ 2 & 73 $\pm$ 10 & 0.016  &    0.21 $\pm$    0.03 &   \\\hline
Sgr C & 12 &H$^{13}$CO$^+$ & -14 $\pm$ 4 & 75 $\pm$ 10 & 0.0063  &    0.09 $\pm$    0.01 &   \\\hline
Sgr C & 13* &CO & 60 $\pm$ 5 & 30 $\pm$ 5 & 1.3 $\pm$ 0.1 &   24 $\pm$    4 & 3 kpc far \\\hline
Sgr C & 13 &SiO & 57 $\pm$ 1 & 29 $\pm$ 4 & 0.007 $\pm$ 0.001 &    0.034 $\pm$    0.007 & 3 kpc far \\\hline
Sgr C & 13 &H$^{13}$CO$^+$ & 57 $\pm$ 1 & 33 $\pm$ 3 & 0.0026 $\pm$ 0.0002 &    0.015 $\pm$    0.002 & 3 kpc far \\\hline
Sgr C & 14 &CO & 126 $\pm$ 6 & 76 $\pm$ 14 & 1.5 $\pm$ 0.2 &   68 $\pm$   16 & EMR \\\hline
Sgr C & 14 &HCO$^+$ & 132 $\pm$ 1 & 85 $\pm$ 1 & 0.057 &    1.57  & EMR \\\hline
Sgr C & 14 &SiO & 153 $\pm$ 1 & 26 $\pm$ 2 & 0.0064  &    0.030  & EMR \\\hline
Sgr D & 15 &CO & -53 $\pm$ 48 & 62 $\pm$ 73 & 2.2 $\pm$ 1.6 &   59 $\pm$   83 &  \\\hline
Sgr D & 15 &HCO$^+$ & -56 $\pm$ 2 & 73 $\pm$ 4 & 0.043  &    1.35      &  \\\hline
Sgr D & 15 &SiO & -48 $\pm$ 3 & 65 $\pm$ 4 & 0.009  &    0.118  &  \\\hline
Sgr D & 16 &CO & 8 $\pm$ 26 & 56 $\pm$ 70 & 3.0 $\pm$ 1.8 &   74 $\pm$  104 &   \\\hline
Sgr D & 16 &HCO$^+$ & -10 $\pm$ 1 & 32 $\pm$ 2 & 0.054  &    0.75 $\pm$    0.07 &   \\\hline
Sgr D & 16 &SiO & 18 $\pm$ 1 & 58 $\pm$ 3 & 0.019  &    0.23  &   \\\hline
Sgr D & 16 &H$^{13}$CO$^+$ & 4 $\pm$ 2 & 79 $\pm$ 5 & 0.0048  &    0.077  &   \\\hline
Sgr D & 17 &CO & 86 $\pm$ 8 & 65 $\pm$ 28 & 5.4 $\pm$ 0.8 &  157 $\pm$   71 & Sgr D cloud \\\hline
Sgr D & 17 &HCO$^+$ & 27 $\pm$ 1 & 31 $\pm$ 2 & 0.062 &    0.83 & Sgr D cloud \\\hline
Sgr D & 17 &HCO$^+$ & 88 $\pm$ 1 & 71 $\pm$ 1 & 0.31  &    9.6  & Sgr D cloud   \\\hline
Sgr D & 17 &SiO & 81 $\pm$ 1 & 55 $\pm$ 1 & 0.085  &    0.941  & Sgr D cloud \\\hline
Sgr D & 17 &H$^{13}$CO$^+$ & 83 $\pm$ 1 & 40 $\pm$ 1 & 0.026  &    0.216  & Sgr D cloud\\\hline
Sgr D & 18 &CO & 176 $\pm$ 26 & 60 $\pm$ 62 & 1.3 $\pm$ 0.8 &   35 $\pm$   42 & EMR  \\\hline
Sgr D & 18* &HCO$^+$ & 180 $\pm$ 5 & 50 $\pm$ 5 & 0.045 $\pm$ 0.005 &    1.0 $\pm$    0.1 & EMR  \\\hline
Sgr E & 19 & CO & -201$\pm$ 14 & 33 $\pm$ 34 & 0.7$\pm$  0.6 &  20.7 $\pm$ 27.9 & Sgr E cloud\\\hline
Sgr E & 19 &HCO$^+$ &-203$\pm$ 1& 29$\pm$ 1 &0.024 & 0.477 & Sgr E cloud\\\hline
Sgr E & 19 &SiO & -180 $\pm$ 3 & 40 $\pm$ 7 & 0.0030 $\pm$ 0.0003 & 0.040 $\pm$0.008& Sgr E cloud\\\hline
Sgr E & 19 &H$^{13}$CO$^+$ &-185 $\pm$ 1 &40 $\pm$ 2 & 0.0032 & 0.042 & Sgr E cloud\\\hline
Sgr E & 20 & CO & -128 $\pm$ 23 & 83 $\pm$ 96  & 0.76$\pm$ 0.37 &  57 $\pm$ 71& EMR  \\\hline
Sgr E & 20 &HCO$^+$ &-124 $\pm$1 & 103 $\pm$ 4 & 0.03  & 2 & EMR  \\\hline
Sgr E & 20 &SiO &-124 $\pm$2 & 40 $\pm$ 4 &0.005   & 0.063 $\pm$0.008& EMR \\\hline
Sgr E & 20 &H$^{13}$CO$^+$ & -123 $\pm$ 1 & 22  $\pm$ 2  & 0.0016$\pm$ 0.0001 &0.012 $\pm$0.002& EMR \\\hline
Sgr E & 21 & CO & -59 $\pm$ 25 & 28$\pm$ 43 &  1.2 $\pm$ 1.6  &  31$\pm$ 62&  \\\hline
Sgr E & 21 & CO & -26 $\pm$ 9 & 33 $\pm$ 44 & 2.7 $\pm$0.6 &  79$\pm$ 108 &  \\\hline
Sgr E & 21 & CO & -3 $\pm$4& 11$\pm$ 11& 2.1$\pm$ 2.2  &  22$\pm$ 32 & \\\hline
Sgr E & 21 &HCO$^+$ & -31 $\pm$ 1& 49 $\pm$ 1 & 0.096 & 3.2&\\\hline
Sgr E & 21 &SiO &-23 $\pm$ 1& 67 $\pm$ 1 &0.021  &0.47&\\\hline
Sgr E & 21 &H$^{13}$CO$^+$ &-23 $\pm$ 1 &  81$\pm$ 1 &0.007&0.19 &\\\hline
Sgr E & 22 & CO & 13 $\pm$ 5 & 16 $\pm$ 11& 2.1 $\pm$0.9 & 30$\pm$ 25 & MM \\\hline
Sgr E & 22* &HCO$^+$ & 12 $\pm$ 5 & 25 $\pm$ 5 & 1.1 0.2 &  & MM \\\hline
Sgr E & 23* & CO & 140 $\pm$ 5 &  55 $\pm$ 5 & 0.50 $\pm$ 0.05 &24.7$\pm$3.3  & EMR \\\hline
Sgr E & 23 & HCO$^+$ &131 $\pm$ 1 & 68 $\pm$ 1 & 0.028 &1.28 & EMR \\\hline
Sgr E & 23 &SiO &134 $\pm$ 1 & 36 $\pm$ 1 &0.009  &0.114 & EMR \\\hline
Sgr E & 23 &H$^{13}$CO$^+$ &105 $\pm$ 1 & 60  $\pm$ 2 & 0.002 &0.048 & EMR \\\hline
1.3 complex & 24 &CO & -1 $\pm$ 8 & 88 $\pm$ 17 & 2.9 $\pm$ 0.2 &  265 $\pm$   55 &  \\\hline
1.3 complex & 24 &HCO$^+$ & -22 $\pm$ 1 & 75 $\pm$ 1 & 0.048  &    2.99  &  \\\hline
1.3 complex & 24 &SiO & -16 $\pm$ 1 & 54 $\pm$ 1 & 0.018  &    0.462  &  \\\hline
1.3 complex & 24 &H$^{13}$CO$^+$ & -24 $\pm$ 1 & 47 $\pm$ 2 & 0.0046  &    0.102  &  \\\hline
1.3 complex & 25 &CO & 85 $\pm$ 6 & 64 $\pm$ 18 & 3.1 $\pm$ 1.0 &  206 $\pm$   89 & 1.3 complex cloud \\\hline
1.3 complex & 25 &HCO$^+$ & 83 $\pm$ 0 & 90 $\pm$ 1 & 0.18  &   13.6  & 1.3 complex cloud \\\hline
1.3 complex & 25 &SiO & 81 $\pm$ 1 & 76 $\pm$ 1 & 0.076  &    2.735  & 1.3 complex cloud \\\hline
1.3 complex & 25 &H$^{13}$CO$^+$ & 75 $\pm$ 1 & 82 $\pm$ 1 & 0.01  &    0.388  & 1.3 complex cloud \\\hline
1.3 complex & 26 &CO & 167 $\pm$ 43 & 121 $\pm$ 73 & 1.1 $\pm$ 0.3 &  143 $\pm$   93 & EMR \\\hline
1.3 complex & 26 &HCO$^+$ & 178 $\pm$ 2 & 86 $\pm$ 3 & 0.031  &    2.19  & EMR \\\hline
M+3.2+0.3 & 27 &CO & -42 $\pm$ 18 & 35 $\pm$ 42 & 0.5 $\pm$ 0.4 &   29 $\pm$   41 & 3 kpc \\\hline
M+3.2+0.3 & 27 &H$^{13}$CO$^+$ & -45 $\pm$ 3 & 48 $\pm$ 7 & 0.0025 $\pm$ 0.0002 &    0.033 $\pm$    0.006 & 3 kpc \\\hline
M+3.2+0.3 & 28 &CO & 11 $\pm$ 4 & 40 $\pm$ 14 & 2.3 $\pm$ 0.5 &  150 $\pm$   62 &   \\\hline
M+3.2+0.3 & 28 &HCO$^+$ & 32 $\pm$ 1 & 84 $\pm$ 2 & 0.09 &    6.4  &   \\\hline
M+3.2+0.3 & 28 &SiO & 23 $\pm$ 1 & 51 $\pm$ 1 & 0.045  &    0.64  &   \\\hline
M+3.2+0.3 & 28 &H$^{13}$CO$^+$ & 24 $\pm$ 2 & 55 $\pm$ 5 & 0.0094  &    0.14  &   \\\hline
M+3.2+0.3 & 29 &CO & 93 $\pm$ 10 & 97 $\pm$ 49 & 1.7 $\pm$ 0.3 &  276 $\pm$  146 & Clump 2 \\\hline
M+3.2+0.3 & 29 &HCO$^+$ & 103 $\pm$ 1 & 68 $\pm$ 3 & 0.074  &    4.4  & Clump 2 \\\hline
M+3.2+0.3 & 29 &SiO & 80 $\pm$ 1 & 54 $\pm$ 2 & 0.039  &    0.59  & Clump 2 \\\hline
M+3.2+0.3 & 29 &H$^{13}$CO$^+$ & 85 $\pm$ 6 & 63 $\pm$ 9 & 0.0042  &    0.07 $\pm$    0.01 & Clump 2 \\\hline
M+3.2+0.3 & 30 &CO & 159 $\pm$ 11 & 27 $\pm$ 37 & 0.6 $\pm$ 0.6 &   24 $\pm$   45 &   \\\hline
M+3.2+0.3 & 30 &HCO$^+$ & 159 $\pm$ 1 & 38 $\pm$ 1 & 0.037  &    1.21  &   \\\hline
M+3.2+0.3 & 30 &SiO & 139 $\pm$ 1 & 61 $\pm$ 2 & 0.031  &    0.52 $\pm$    0.02 &   \\\hline
M+3.2+0.3 & 31 &CO & 237 $\pm$ 32 & 55 $\pm$ 76 & 0.3 $\pm$ 0.3 &   25 $\pm$   45 & EMR \\\hline
M+3.2+0.3 & 31 &HCO$^+$ & 219 $\pm$ 13 & 137 $\pm$ 24 & 0.0055  &    0.6 $\pm$    0.1 & EMR  \\\hline
M$-$5.3+0.4 & 32 &CO & -81 $\pm$ 3 & 14 $\pm$ 7 & 1.0 $\pm$ 0.4 &   12 $\pm$    8 & 3 kpc \\\hline
M$-$5.3+0.4 & 32 &HCO$^+$ & -78 $\pm$ 4 & 29 $\pm$ 7 & 0.012 $\pm$ 0.002 &    0.19 $\pm$    0.06 & 3 kpc \\\hline
M$-$5.3+0.4 & 32 &SiO & -91 $\pm$ 1 & 8 $\pm$ 3 & 0.0027 $\pm$ 0.0008 &    0.006 $\pm$    0.003 & 3 kpc \\\hline
M$-$5.3+0.4 & 32 &H$^{13}$CO$^+$ & -80 $\pm$ 1 & 8 $\pm$ 3 & 0.0025 $\pm$ 0.0008 &    0.005 $\pm$ 0.003 & 3 kpc \\\hline
M$-$5.3+0.4 & 33 &CO & -44 $\pm$ 8 & 25 $\pm$ 21 & 0.7 $\pm$ 0.3 &   16 $\pm$   15 & Norma \\\hline
M$-$5.3+0.4 & 33 &HCO$^+$ & -43 $\pm$ 5 & 33 $\pm$ 23 & 0.008 $\pm$ 0.001 &    0.15 $\pm$    0.1 & Norma \\\hline
M$-$5.3+0.4 & 33* &SiO & -50 $\pm$ 5 & 20 $\pm$ 5 & 0.008  &    0.04 $\pm$    0.01 & Norma \\\hline
M$-$5.3+0.4 & 33 &H$^{13}$CO$^+$ & -55 $\pm$ 4 & 33 $\pm$ 16 & 0.0019 $\pm$ 0.0004 &    0.018 $\pm$    0.009 & Norma \\\hline
M$-$5.3+0.4 & 34 &CO & -18 $\pm$ 3 & 17 $\pm$ 7 & 1.6 $\pm$ 0.4 &   24 $\pm$   12 & Crux \\\hline
M$-$5.3+0.4 & 34 &HCO$^+$ & -21 $\pm$ 1 & 9 $\pm$ 2 & 0.021 $\pm$ 0.003 &    0.11 $\pm$    0.03 & Crux \\\hline
M$-$5.3+0.4 & 34*&SiO & -20 $\pm$ 5 & 10 $\pm$ 5 & 0.0078  &    0.02 $\pm$    0.01 & Crux \\\hline
M$-$5.3+0.4 & 34 &H$^{13}$CO$^+$ & -23 $\pm$ 2 & 17 $\pm$ 4 & 0.0037 $\pm$ 0.0006 &    0.017 $\pm$    0.005 & Crux \\\hline
M$-$5.3+0.4 & 35 &CO & 3 $\pm$ 1 & 10 $\pm$ 2 & 3.0 $\pm$ 0.5 &   27 $\pm$    7 & MM \\\hline
M$-$5.3+0.4 & 35* &HCO$^+$ & 5 $\pm$ 5 & 10 $\pm$ 5 & 0.030 $\pm$ 0.003 &    0.18 $\pm$    0.09 & MM \\\hline
M$-$5.3+0.4 & 35 &SiO & 12 $\pm$ 3 & 37 $\pm$ 9 & 0.004 $\pm$ 0.0004 &    0.04 $\pm$    0.01 & MM \\\hline
M$-$5.3+0.4 & 35 &H$^{13}$CO$^+$ & 11 $\pm$ 2 & 26 $\pm$ 7 & 0.0024 $\pm$ 0.0004 &    0.017 $\pm$    0.005 & MM \\\hline
M$-$5.3+0.4 & 36 &CO & 87 $\pm$ 3 & 42 $\pm$ 7 & 1.7 $\pm$ 0.2 &   65 $\pm$   13 & Clump 1 \\\hline
M$-$5.3+0.4 & 36 &HCO$^+$ & 86 $\pm$ 1 & 41 $\pm$ 1 & 0.069  &    1.65  & Clump 1 \\\hline
M$-$5.3+0.4 & 36 &SiO & 72 $\pm$ 1 & 51 $\pm$ 2 & 0.015  &    0.206  & Clump 1 \\\hline
M$-$5.3+0.4 & 36 &H$^{13}$CO$^+$ & 83 $\pm$ 2 & 92 $\pm$ 5 & 0.0061  &    0.15 $\pm$    0.01 & Clump 1 \\\hline
M$-$4.4+0.6 & 37 &CO & -75 $\pm$ 1 & 28 $\pm$ 3 & 1.2 $\pm$ 0.1 &    8 $\pm$    1 & 3 kpc \\\hline
M$-$4.4+0.6 & 37* &HCO$^+$ & -70 $\pm$ 5 & 10 $\pm$ 5 & 0.014 $\pm$ 0.003 &    0.025 $\pm$    0.01 & 3 kpc \\\hline
M$-$4.4+0.6 & 37 &SiO & -75 $\pm$ 1 & 14 $\pm$ 2 & 0.0062 $\pm$ 0.0009 &    0.009 $\pm$    0.002 & 3 kpc \\\hline
M$-$4.4+0.6 & 37 &H$^{13}$CO$^+$ & -70 $\pm$ 2 & 22 $\pm$ 4 & 0.0047 $\pm$ 0.0006 &    0.011 $\pm$    0.002 & 3 kpc \\\hline
M$-$4.4+0.6 & 38 &CO & -34 $\pm$ 1 & 16 $\pm$ 2 & 1.3 $\pm$ 0.1 &    5 $\pm$    0.9 & Norma \\\hline
M$-$4.4+0.6 & 38* &HCO$^+$ & -45 $\pm$ 5 & 20 $\pm$ 5 & 0.018 $\pm$ 0.003 &    0.06 $\pm$    0.02 & Norma \\\hline
M$-$4.4+0.6 & 38 &SiO & -40 $\pm$ 5 & 15 $\pm$ 5 & 0.008  &    0.012 $\pm$    0.004 & Norma \\\hline
M$-$4.4+0.6 & 38 &H$^{13}$CO$^+$ & -37 $\pm$ 1 & 10 $\pm$ 2 & 0.006 $\pm$ 0.001 &    0.006 $\pm$    0.002 & Norma \\\hline
M$-$4.4+0.6 & 39* &CO & -20 $\pm$ 5 & 10 $\pm$ 5 & 0.72 $\pm$ 0.05 &    1.8 $\pm$    0.9 & Crux \\\hline
M$-$4.4+0.6 & 39* &HCO$^+$ & -30 $\pm$ 5 & 10 $\pm$ 5 & 0.036 $\pm$ 0.003 &    0.06 $\pm$    0.03 & Crux \\\hline
M$-$4.4+0.6 & 39 &SiO & -20 $\pm$ 5 & 20 $\pm$ 5 & 0.0092  &    0.018 $\pm$    0.005 & Crux \\\hline
M$-$4.4+0.6 & 39 &H$^{13}$CO$^+$ & -23 $\pm$ 1 & 6 $\pm$ 2 & 0.005 $\pm$ 0.001 &    0.003 $\pm$    0.001 & Crux \\\hline
M$-$4.4+0.6 & 40* &CO & 0 $\pm$ 5 & 10 $\pm$ 5 & 2.75  &    7 $\pm$    3 & MM \\\hline
M$-$4.4+0.6 & 40 &H$^{13}$CO$^+$ & -1 $\pm$ 2 & 27 $\pm$ 6 & 0.0041 $\pm$ 0.0006 &    0.012 $\pm$    0.003 & MM \\\hline
M$-$4.4+0.6 & 41* &CO & 18 $\pm$ 5 & 20 $\pm$ 5 & 1.25 $\pm$ 0.05 &    6 $\pm$    2 &   \\\hline
M$-$4.4+0.6 & 41 &SiO & 25 $\pm$ 2 & 30 $\pm$ 4 & 0.011 $\pm$ 0.001 &    0.033 $\pm$    0.005 &   \\\hline
M$-$4.4+0.6 & 42 &CO & 70 $\pm$ 2 & 55 $\pm$ 6 & 1.55 $\pm$ 0.08 &   21 $\pm$    2 & M$-$4.4+0.6 cloud \\\hline
M$-$4.4+0.6 & 42 &HCO$^+$ & 71 $\pm$ 1 & 49 $\pm$ 1 & 0.048 &    0.42  & M$-$4.4+0.6 cloud \\\hline
M$-$4.4+0.6 & 42 &SiO & 72 $\pm$ 1 & 51 $\pm$ 3 & 0.014  &    0.073  & M$-$4.4+0.6 cloud \\\hline
M$-$4.4+0.6 & 42 &H$^{13}$CO$^+$ & 65 $\pm$ 1 & 39 $\pm$ 3 & 0.0073  & 0.031 $\pm$   0.003 & M$-$4.4+0.6 cloud \\\hline
M$-$3.8+0.9 & 43 &CO & -67 $\pm$ 6 & 81 $\pm$ 14 & 1.0 $\pm$ 0.1 &   28 $\pm$    5 & M$-$3.8+0.9 cloud \\\hline
M$-$3.8+0.9 & 43 &HCO$^+$ & -79 $\pm$ 1 & 61 $\pm$ 1 & 0.067  &    1.05  & M$-$3.8+0.9 cloud \\\hline
M$-$3.8+0.9 & 43 &SiO & -80 $\pm$ 2 & 79 $\pm$ 4 & 0.014 &    0.2  & M$-$3.8+0.9 cloud \\\hline
M$-$3.8+0.9 & 43 &H$^{13}$CO$^+$ & -60 $\pm$ 4 & 73 $\pm$ 7 & 0.0097 &    0.13 $\pm$    0.01 & M$-$3.8+0.9 cloud \\\hline
M$-$3.8+0.9 & 44 &CO & 0 $\pm$ 2 & 37 $\pm$ 5 & 2.1 $\pm$ 0.2 &   26 $\pm$    4 & MM \\\hline
M$-$3.8+0.9 & 44 &HCO$^+$ & -4 $\pm$ 1 & 55 $\pm$ 2 & 0.058 &    0.83  & MM \\\hline
M$-$3.8+0.9 & 44 &SiO & -5 $\pm$ 1 & 43 $\pm$ 3 & 0.014  &    0.109  & MM \\\hline
M$-$3.8+0.9 & 44 &H$^{13}$CO$^+$ & 5 $\pm$ 2 & 54 $\pm$ 4 & 0.012  &    0.12 $\pm$    0.01 & MM \\\hline
M$-$3.8+0.9 & 45 &CO & 60 $\pm$ 13 & 53 $\pm$ 51 & 0.3 $\pm$ 0.1 &    5 $\pm$    5 & 3 kpc far \\\hline
M$-$3.8+0.9 & 45*&HCO$^+$ & 60 $\pm$ 5 & 15 $\pm$ 5 & 0.013 $\pm$ 0.002 & 0.05  $\pm$ 0.02    & 3 kpc far \\\hline
M$-$3.8+0.9 & 45 &SiO & 40 $\pm$ 2 & 16 $\pm$ 5 & 0.004 $\pm$ 0.001 &    0.013 $\pm$    0.005 & 3 kpc far \\\hline
M$-$3.8+0.9 & 45 &H$^{13}$CO$^+$ & 60 $\pm$ 1 & 8 $\pm$ 2 & 0.004 $\pm$ 0.001 &  0.005 $\pm$    0.002 & 3 kpc far \\\hline
M$-$3.8+0.9 & 46 &CO & 106 $\pm$ 3 & 19 $\pm$ 7 & 0.74 $\pm$ 0.20 &    5 $\pm$    2 & 135 km/s \\\hline
M$-$3.8+0.9 & 46 &HCO$^+$ & 108 $\pm$ 1 & 17 $\pm$ 2 & 0.019 $\pm$ 0.003 &    0.08 $\pm$    0.02 & 135 km/s \\\hline
M$-$3.8+0.9 & 46* &SiO & 80 $\pm$ 5 & 20 $\pm$ 5 & 0.012  &   0.04  $\pm$ 0.01   & 135 km/s \\\hline
M$-$3.8+0.9 & 46 &H$^{13}$CO$^+$ & 95 $\pm$ 1 & 7 $\pm$ 1 & 0.006 $\pm$ 0.001 &    0.008 $\pm$   0.002 & 135 km/s \\\hline
M+5.3$-$0.3 & 47 &CO & -27 $\pm$ 1 & 9 $\pm$ 2 & 1.4 $\pm$ 0.2 &    9 $\pm$    2 & 3 kpc \\\hline
M+5.3$-$0.3 & 47 &HCO$^+$ & -28 $\pm$ 1 & 15 $\pm$ 2 & 0.013 $\pm$ 0.001 &    0.06 $\pm$    0.01 & 3 kpc \\\hline
M+5.3$-$0.3 & 48 &CO & 9 $\pm$ 1 & 12 $\pm$ 2 & 2.9 $\pm$ 0.2 &   25 $\pm$    4 & MM \\\hline
M+5.3$-$0.3 & 48 &CO & 23 $\pm$ 3 & 11 $\pm$ 6 & 0.7 $\pm$ 0.2 &    6 $\pm$    4 & MM \\\hline
M+5.3$-$0.3 & 48 &HCO$^+$ & 23 $\pm$ 2 & 39 $\pm$ 3 & 0.05  &    0.61  & MM \\\hline
M+5.3$-$0.3 & 48* &SiO & 26 $\pm$ 1 & 48 $\pm$ 1 & 0.014  &    0.207  & MM \\\hline
M+5.3$-$0.3 & 48 &H$^{13}$CO$^+$ & 8 $\pm$ 13 & 49 $\pm$ 14 & 0.003 $\pm$ 0.001 &    0.043 $\pm$    0.02 & MM \\\hline
M+5.3$-$0.3 & 49 &CO & 40 $\pm$ 12 & 62 $\pm$ 15 & 1.3 $\pm$ 0.5 &   55 $\pm$   26 &   \\\hline
M+5.3$-$0.3 & 49 &HCO$^+$ & 59 $\pm$ 1 & 32 $\pm$ 3 & 0.062 $\pm$ 0.007 &    0.6 $\pm$    0.1 &   \\\hline
M+5.3$-$0.3 & 49* &SiO & 65 $\pm$ 5 & 35 $\pm$ 5 & 0.017  &    0.18 $\pm$    0.03 &   \\\hline
M+5.3$-$0.3 & 49 &H$^{13}$CO$^+$ & 54 $\pm$ 5 & 51 $\pm$ 11 & 0.0068 $\pm$ 0.0009 &    0.11 $\pm$    0.03 &   \\\hline
M+5.3$-$0.3 & 50 &CO & 104 $\pm$ 28 & 83 $\pm$ 52 & 0.8 $\pm$ 0.2 &   45 $\pm$   31 & M+5.3$-$0.3 cloud \\\hline
M+5.3$-$0.3 & 50 &HCO$^+$ & 98 $\pm$ 2 & 47 $\pm$ 8 & 0.061 $\pm$ 0.007 &  0.81   $\pm$  0.13 & M+5.3$-$0.3 cloud \\\hline
M+5.3$-$0.3 & 50* &SiO & 105 $\pm$ 5 & 60 $\pm$ 5 & 0.021  &    0.38 $\pm$    0.03 & M+5.3$-$0.3 cloud \\\hline
M+5.3$-$0.3 & 50 &H$^{13}$CO$^+$ & 120 $\pm$ 4 & 61 $\pm$ 9 & 0.0045  &    0.08 $\pm$    0.01 & M+5.3$-$0.3 cloud \\\hline
M+5.3$-$0.3 & 51 &CO & 190 $\pm$ 6 & 38 $\pm$ 15 & 0.4 $\pm$ 0.1 &   10 $\pm$    5 & 135 km/s \\\hline
M+5.3$-$0.3 & 51 &HCO$^+$ & 150 $\pm$ 6 & 50 $\pm$ 8 & 0.020 $\pm$ 0.002 &  0.31   $\pm$ 0.06    &135 km/s  \\\hline
M+5.3$-$0.3 & 51 &H$^{13}$CO$^+$ & 173 $\pm$ 1 & 12 $\pm$ 2 & 0.0041 $\pm$0.0004 &    0.015 $\pm$    0.003 & 135 km/s \\\hline
\end{longtable}
%\end{longtable}
\begin{list}{}{}
\item In most, the nominal Gaussian fits do the data give uncertainties smaller that $10\%$, the
estimated relative calibration uncertainty of our data, only in cases where the
errors from the Gaussian fit exceed $10\%$, these uncertainties are indicated.
\item The size of each region is given in Table \ref{table1}.
\item $^{\mathrm{a}}$: Gaussian FWHM.
\item Central velocity, Velocity width and temperature peaks (T$_0$) derived by the
  average spectrum shown in Appendix D.
\item EMR: ``Expanding Molecular Ring'' \citep{Bitran_1987}
\item MM: ``Main Maximum'' \citep{Bitran_1987}
\item 3 kpc: ``3 kpc Arm'' (\citealp{Bitran_1987, Sawada_et_al_2001})
\item Norma: ``Norma Arm'' (\citealp{Bitran_1987, Sawada_et_al_2001})
\item Crux: ``Crux Arm'' (\citealp{Bitran_1987, Sawada_et_al_2001})
\item 135 km/s: ``135 km s$^{-1}$ Arm'' \citealp{Bitran_1987,Bania_1980}
\item Clump 1 (\citealp{Bania_1977,Bania_et_al_1986,Bitran_1987})
\item Clump 2 (\citealp{Bania_1977,Bitran_1987})
\item The cloud number 15 is formed by the 3 kpc, Crux and Norma arms.
\item M$-$4.4+0.6 cloud, identifyed by \citet{Bitran_1987}.
\item M$-$3.8+0.9 cloud, identifyed by \citet{Bitran_1987}.
\item M+5.3$-$0.3 cloud, identifyed by \citet{Bitran_1987}.
\end{list}
}
}
\end{document}